\documentclass{aa}
\usepackage{psfig}
\usepackage{natbib}
\bibpunct{(}{)}{,}{a}{}{,} 
\usepackage{graphicx}

\def\la{\;
\raise0.3ex\hbox{$<$\kern-0.75em\raise-1.1ex\hbox{$\sim$}}\; }
\def\ga{\;
\raise0.3ex\hbox{$>$\kern-0.75em\raise-1.1ex\hbox{$\sim$}}\; }

\newcommand{\dmm}{$\Delta\mu/\mu$}

\newcommand{\kms}{km~s$^{-1}$}
\newcommand{\cms}{cm~s$^{-1}$}
\newcommand{\ms}{m~s$^{-1}$}

\newcommand{\cmm}{cm$^{-3}$}
\newcommand{\etal}{{et al.}}
\newcommand{\nhhh}{NH$_3$}
\newcommand{\nnhp}{N$_2$H$^+$}
\newcommand{\nndp}{N$_2$D$^+$}
\newcommand{\hcccn}{HC$_3$N}
\newcommand{\DV}{$\Delta V$}

\begin{document}

\title{
Searching for chameleon-like scalar fields with the ammonia method.\thanks{Based 
on observations obtained with the 100-m telescope at Effelsberg/Germany which is operated by 
the Max-Planck Institut f\"ur Radioastronomie on behalf of the Max-Planck-Gesellschaft (MPG).
} 
}
\subtitle{II. Mapping of cold molecular cores in \nhhh\ and \hcccn\ lines}
\author{S. A. Levshakov\inst{1,2,3}
\and
A. V. Lapinov\inst{4}
\and
C. Henkel\inst{5}
\and
P. Molaro\inst{1}
\and
D. Reimers\inst{6}
\and
M. G. Kozlov\inst{7}
\and
I. I. Agafonova\inst{3}
}
\institute{
INAF-Osservatorio Astronomico di Trieste, Via G. B. Tiepolo 11,
34131 Trieste, Italy\\
\email{lev@astro.ioffe.rssi.ru} 
\and
Key Laboratory for Research in Galaxies and Cosmology,
Shanghai Astronomical Observatory, CAS, 80 Nandan Road, Shanghai 200030,
P. R. China
\and
Ioffe Physical-Technical Institute, 
Polytekhnicheskaya Str. 26, 194021 St.~Petersburg, Russia
\and
Institute for Applied Physics, Uljanov Str. 46, 603950 Nizhny Novgorod, Russia
\and
Max-Planck-Institut f\"ur Radioastronomie, Auf dem H\"ugel 69, D-53121 Bonn, Germany
\and
Hamburger Sternwarte, Universit\"at Hamburg,
Gojenbergsweg 112, D-21029 Hamburg, Germany
\and
Petersburg Nuclear Physics Institute, 188300 Gatchina, Russia
}
\date{Received 00  ; Accepted 00}
\abstract
{In our previous work we found a statistically significant 
offset \DV\ $\approx 27$ \ms\ between the radial velocities of the 
\hcccn\ $J = 2-1$ and \nhhh\ $(J,K) = (1,1)$ transitions
observed in molecular cores from the Milky Way.
This may indicate that the electron-to-proton mass ratio,
$\mu \equiv m_{\rm e}/m_{\rm p}$, 
increases by $\sim 3\times10^{-8}$
when measured under interstellar conditions with matter densities
of more than 10 orders of magnitude lower as compared with
laboratory (terrestrial) environments. 
}
{
We map four molecular cores L1498, L1512, L1517, and L1400K 
selected from our previous sample in order to
estimate systematic effects in \DV\ due to possible velocity gradients 
or other sources across the cloud, 
and to check the reproducibility of the velocity offsets 
on the year-to-year time base.
}
{
We use the ammonia method, which involves observations of inversion lines of \nhhh\ complemented by
rotational lines of other molecular species and allows us to test changes in $\mu$ due to a 
higher sensitivity of the inversion frequencies to the $\mu$-variation as compared with the
rotational frequencies.
} 
{
We find that in two cores L1498 and L1512 
the \nhhh\ (1,1) and \hcccn\ (2--1) 
transitions closely trace the same material
and show an offset of
$\Delta V  \equiv  V_{lsr}$(\hcccn) -- $V_{lsr}$(\nhhh) =
$26.9\pm1.2_{\rm stat}\pm3.0_{\rm sys}$ \ms\
throughout the entire clouds.
The offsets measured in L1517B and L1400K are
$46.9\pm3.3_{\rm stat}\pm3.0_{\rm sys}$ \ms, and
$8.5\pm3.4_{\rm stat}\pm3.0_{\rm sys}$ \ms, respectively, and are,
probably, subject to Doppler shifts due to spatial 
segregation of \hcccn\ versus \nhhh.
We also determine
frequency shifts caused by external electric and magnetic fields,
and by the cosmic black body radiation-induced Stark effect
and find that they are less than 1 \ms.
}
{
The measured velocity offset in L1498 and L1512, being expressed in terms of
\dmm\ $\equiv (\mu_{\rm obs} - \mu_{\rm lab})/\mu_{\rm lab}$,
gives
\dmm\ = $(26\pm1_{\rm stat}\pm3_{\rm sys})\times10^{-9}$.
Although this estimate is  based on a limited number of sources and molecular pairs 
used in the ammonia method, it demonstrates a high sensitivity
of radio observations in testing fundamental physics.
The non-zero signal
in \dmm\ should  be further examined as larger and
more 
accurate data sets become available. 
}
\authorrunning{S. A. Levshakov et al.}
\titlerunning{Searching for chameleon-like scalar fields with the ammonia method}

\maketitle

\section{Introduction}
\label{sect-1}

The present paper continues our study 
(Levshakov \etal\ 2008; Molaro \etal\ 2009; Levshakov \etal\ 2010a, hereafter Paper~I) on
differential measurements of the electron-to-proton mass ratio,
$\mu = m_{\rm e}/m_{\rm p}$, 
by means of high resolution spectral observations (FWHM $\sim 30-40$ \ms) of
narrow emission lines (FWHM $< 200$ \ms) of N-bearing molecules arising in
cold and dense molecular clouds ($T_{\rm kin} \sim 10$ K, $n \sim 10^4 - 10^5$ \cmm).
This study is aimed, in general, at testing the Einstein equivalence principle of local position invariance
(LPI) which states that outcomes of physical non-gravitational experiments should be independent
of their position in space-time (e.g., Dent 2008).
However, the violation of LPI is anticipated in non-standard physical theories
associated with varying fundamental constants and, in particular, in 
those dealing with dark energy.
A concept of dark energy with negative pressure appeared in physics
long before the discovery of the accelerating universe 
through observations of nearby and distant supernovae type Ia
(Perlmutter \etal\ 1998; Riess \etal\ 1998). 
Early examples of dark energy in a form of a scalar field with a self-interaction potential
can be found in a review by Peebles \& Ratra (2003).
To explain the nature of dark energy, many sophisticated models have been suggested
(see, e.g., Caldwell \etal\ 1998),
and among them the scalar fields which are ultra-light in cosmic vacuum but possess
an effectively large mass locally when they are coupled to
ordinary matter by the so-called chameleon mechanism 
(Khoury \& Weltman 2004a,b; Brax \etal\ 2004; Avelino 2008; 
Burrage \etal\ 2009; Davis \etal\ 2009; Brax 2009; Upadhye \etal\ 2010; Brax \& Zioutas 2010).
A subclass of such models considered by Olive \& Pospelov (2008) predicts that
fundamental physical quantities such as elementary particle masses 
and low-energy coupling constants may also depend on the local matter density.

In the Standard Model of particle physics 
there are two fundamental parameters with the dimension
of mass~-- the Higgs vacuum expectation value (VEV $\sim 200$ GeV), which determines the electroweak
unification scale, and the quantum chromodynamics (QSD) scale $\Lambda_{\rm QCD} \sim 220$ MeV,
which characterizes the strength of the strong interaction.
The masses of the elementary particles such as the electron $m_{\rm e}$ and 
the quarks $m_{\rm q}$ are proportional to the Higgs VEV,
and the mass of the composite particle proton $m_{\rm p}$ is proportional to $\Lambda_{\rm QCD}$.
Therefore, by probing the dimensionless mass ratio $\mu = m_{\rm e}/m_{\rm p}$ we can
experimentally test of the ratio of the electroweak scale to the strong scale 
at different physical conditions (Chin \etal\ 2009). 

In this regard, the masses of the elementary particles mediated by the chameleon-like scalar fields 
should depend on the local density of baryons and electrons. Among the observationally accessible
media the most dramatic difference of local densities occurs between terrestrial laboratories
and cosmic space. For instance, a typical interstellar molecular cloud with the kinetic temperature
$T_{\rm kin} \sim 10$ K and the gas number density 
$n_{{\scriptscriptstyle \rm H}_2} \sim 10^{5}$ \cmm\ has the local density of  
$\rho_{\rm space} \approx 3\times10^{-19}$ g~\cmm.
Laboratory vacuum spectrographs usually operate in the pressure range from 1 to 10 mtorr 
and at the temperature regime of $T_{\rm kin} \sim 10-40$ K 
(e.g., Willey \etal\ 2002; Ronningen \& De Lucia 2005; Ross \& Willey 2005).
This gives $\rho_{\rm lab} \approx 3\times10^{-9}$ g~\cmm\ if 
$T_{\rm kin} = 10$ K, $P = 1$ mtorr, and the vacuum chamber is filled by H$_2$.
In this case, the ratio $\rho_{\rm lab}/\rho_{\rm space}$ exceeds $10^{10}$.
Besides, the size of the vacuum chamber is $\sim 10$ cm, and 
its walls are naturally made of extremely heavy materials, i.e.
the chameleon fields in the chamber are subject to high density boundary conditions.

\begin{table*}[t!]
\centering
\caption{Source positions, offsets $(\Delta\alpha,\Delta\delta)$, and date of observations.
}
\label{tbl-1}
\begin{tabular}{lccc r@{,}l ccccc}
\hline
\hline
\noalign{\smallskip}
\multicolumn{1}{c}{Core} & R.A. & Dec.  & $V_{lsr}$ & 
\multicolumn{2}{c}{Offset } & \multicolumn{1}{c}{rms (mK)} & 
\multicolumn{1}{c}{rms (mK)} & Date & UT\\
\multicolumn{1}{c}{} & \multicolumn{2}{c}{(J2000.0)} & (\kms) 
& \multicolumn{2}{c}{(arcsec)} & 23~GHz & 18~GHz & (2010)& (h:m) & \#\\
\noalign{\smallskip}
\hline 
\noalign{\smallskip}
L1498    & 04:10:51.4 & $+$25:09:58 & $+7.8$ & 0&0 & 31 & 18 & Jan 18 & 15:40& 1\\[-2pt]
         &            &             &        & 40&0& 28 & 15 & Jan 18 & 16:00& 2\\[-2pt]
         &            &             &        & 20&0& 27 & 15 & Jan 18 & 18:30& 3\\[-2pt]
         &            &             &        & --40&0& 29 & 16 & Jan 18 & 20:25& 4\\[-2pt]
         &            &             &        & 20&20& 29 & 15 & Jan 18 & 21:45& 5\\[-2pt]
         &            &             &        & 20&--20& 29 & 18 & Jan 18 & 23:00& 6\\[-2pt]
         &            &             &        & 80&0& 32 & 17 & Jan 19 & 00:20& 7\\[-2pt]
         &            &             &        & 20&0& 26 & 15 & Jan 21 & 21:20& 8\\
L1512   & 05:04:09.6 & $+$32:43:09 & $+7.1$ & 0&0 & 42 & 19 & Jan 19 & 15:00& 1\\[-2pt]
         &            &             &       & 20&--40& 29 & 18 & Jan 19 & 16:20& 2\\[-2pt]
         &            &             &       & --20&40& 28 & 18 & Jan 19 & 17:30& 3\\[-2pt]
         &            &             &       & --40&80& 27 & 16 & Jan 19 & 18:45& 4\\[-2pt]
         &            &             &       & --30&60& 27 & 17 & Jan 19 & 19:50& 5\\[-2pt]
         &            &             &       & --10&20& 26 & 17 & Jan 19 & 21:15& 6\\[-2pt]
         &            &             &       & 10&--20& 27 & 18 & Jan 19 & 22:20& 7\\[-2pt]
         &            &             &       & --40&0& 26 & 17 & Jan 19 & 23:30& 8\\[-2pt]
         &            &             &       & 0&40& 31 & 17 & Jan 20 & 01:30& 9\\[-2pt]
         &            &             &       & 0&0& 27 & 16 & Jan 20 & 23:50& 10\\[-2pt]
         &            &             &       & --50&100& 28 & 14 & Jan 21 & 01:10& 11\\[-2pt]
         &            &             &       & 0&0& 27 & 16 & Jan 21 & 22:30& 12\\[-2pt]
         &            &             &       & 0&0& 29 & 16 & Jan 21 & 23:40& 13\\
L1517BC  & 04:55:17.5 & $+$30:37:49 & $+5.8$& 0&0 & 28 & 15 & Jan 20 & 15:30& 1\\[-2pt]
         &            &             &       & 15&15& 23 & 15 & Jan 20 & 16:50& 2\\[-2pt]
         &            &             &       & --15&--15& 24 & 15 & Jan 20 & 18:00& 3\\[-2pt]
         &            &             &       & 15&--15& 23 & 15 & Jan 20 & 19:20& 4\\[-2pt]
         &            &             &       & --15&15& 24 & 14 & Jan 20 & 20:25& 5\\[-2pt]
         &            &             &       & 0&0& 31 & 12 & Jan 22 & 01:15& 6\\
L1400K   & 04:30:52.0 & $+$54:51:55 & $+3.3$& 0&0 & 23 & 18 & Jan 19 & 02:20& 1\\[-2pt]
         &            &             &       & 0&0& 30 & 16 & Jan 21 & 14:50& 2\\[-2pt]
         &            &             &       & 0&--40& 28 & 16 & Jan 21 & 16:10& 3\\[-2pt]
         &            &             &       & 40&40& 28 & 15 & Jan 21 & 17:25& 4\\

\noalign{\smallskip}
\hline
\noalign{\smallskip}
\end{tabular}
\end{table*}

It follows from the above considerations that the ratio $\mu = m_{\rm e}/m_{\rm p}$
measured in low density environments of the interstellar and/or intergalactic medium
may differ from its terrestrial (laboratory) value. 
To estimate the fractional changes in $\mu$, 
$\Delta\mu/\mu \equiv (\mu_{\rm obs} - \mu_{\rm lab})/\mu_{\rm lab}$,
the so-called ammonia method was elaborated by Flambaum \& Kozlov (2007, hereafter FK)
which involves observations of inversion lines of \nhhh\
complemented by rotational lines of other molecular species.
The inversion transitions of ammonia and its isotopologues
strongly depend on $\mu$ (van Veldhoven \etal\ 2004; FK; Kozlov \etal\ 2010).
For \nhhh, the fractional change in frequency is
$\Delta\nu_{\rm inv}/\nu_{\rm inv} = 4.46$\dmm, whereas 
for rotational lines,
$\Delta\nu_{\rm rot}/\nu_{\rm rot} = \Delta\mu/\mu$. 
Therefore, the
comparison of radial velocities of \nhhh\ inversion lines, $V_{\rm inv}$, 
with radial velocities of rotational transitions, $V_{\rm rot}$, 
provides a sensitive limit on the variation of $\mu$ (FK):
\begin{equation}
{\Delta \mu}/{\mu} \approx 0.289(V_{\rm rot} - V_{\rm inv})/c
\approx 0.3\Delta V/c\ ,
\label{eq1}
\end{equation}
where $c$ is the speed of light.
Thus, measuring the line shifts in cosmic objects where the local matter density is
significantly lower than in laboratory experiments one can probe whether
the mass-ratio $\mu$ is position-invariant or not.  
Such tests will complement current cosmological probes of dark energy
and its couplings to matter.

\nhhh\ was detected in numerous molecular clouds within the Milky Way
(e.g., Jijina \etal\ 1999) and even at significant redshifts (Henkel \etal\ 2005, 2008).
In particular, observations of two molecular absorption-line systems at redshift $z = 0.68$ and $z = 0.89$
towards, respectively, the gravitationally lensed quasars B0218+357 and PKS 1830--211
yielded constraints on \dmm\ down to 1 ppm (FK; Murphy \etal\ 2008;
Menten \etal\ 2008; Henkel \etal\ 2009)\footnote{Hereafter, 
1 ppm = $10^{-6}$, and 1 ppb = $10^{-9}$.}.
However, \dmm\ can be probed at orders of magnitude more sensitive levels
if nearby Galactic clouds are studied. 
Then a spectral resolution of FWHM $\sim 30$ \ms\
allows us to measure the line position with uncertainties of a few \ms.
In 2008-2009, we observed 41 molecular cores in the disk of the Milky Way
in three molecular transitions NH$_3$ $(J,K) = (1,1)$, HC$_3$N $J = 2-1$, and
N$_2$H$^+$ $J = 1-0$ at the radio telescopes in 
Medicina (32-m), Nobeyama (45-m), and Effelsberg (100-m) and obtained
a statistically significant positive velocity offset
between the rotational \hcccn\ (2--1), N$_2$H$^+$ $J = 1-0$ and inversion \nhhh\ (1,1) lines.  
Since these transitions show a good correlation in their spatial distributions, 
such an offset might indicate that 
at low interstellar densities $\mu$ increases compared with its terrestrial value (Paper~I).
The most accurate estimate was obtained from the Effelsberg dataset
consisting of twelve pairs of \nhhh\ and \hcccn\ lines:
\DV\ = $23\pm4_{\rm stat}\pm3_{\rm sys}$ \ms\ which translates into
$\Delta\mu/\mu = 22\pm4_{\rm stat}\pm3_{\rm sys}$ ppb. 
We note that because of a rounded value for the fast Fourier transform spectrometer (FFTS)
channel separation ($\Delta\nu = 1.220$ kHz for \nhhh\ and 
$\Delta\nu = 1.221$ kHz for \hcccn\ instead of 1.220703125 kHz)
not taken into account in Paper~I, the corrected \dmm\ is to be
$26\pm4_{\rm stat}\pm3_{\rm sys}$ ppb.

These observations were carried out in a single position mode, 
i.e. we observed only the central parts of molecular cores
showing strongest \nhhh\ emission. The aim was to find suitable targets for 
most precise measurements of the velocity offsets \DV\ between rotational and inversion transitions.
In the present paper, we report on new observations where 
we measure \DV\ at different positions
across individual clouds in order to test the reproducibility of the velocity offsets
in the presence of the large-scale velocity gradients.
For this purpose, 
from the list of molecular cores observed in our previous study with the Effelsberg telescope,
we selected the objects which comply with the following criteria:
($i$) profiles of the \nhhh\ (1,1) and \hcccn\ (2--1) hyperfine structure (hfs) transitions 
are symmetric and well-described by a single-component Gaussian model, and 
($ii$) the line widths are thermally dominated, i.e.,  
the parameter $\beta = \sigma_v$(\nhhh)$/\sigma_v$(\hcccn) $\geq 1$,
where $\sigma_v$ is the velocity dispersion.
The chosen targets are molecular cores L1498, L1512, L1517B and L1400K which
were already extensively studied in many molecular lines (e.g. Benson \& Myers 1989;
Lee \etal\ 2001; Lee \etal\ 2003; Tafalla \etal\ 2004, 2006; Crapsi \etal\ 2005).   
In the present paper we describe the high-precision measurements of the radial velocities
of \nhhh\ (1,1) and \hcccn\ (2--1) hyperfine transitions in these clouds 
from 31 lines of sight in total.

\begin{figure*}[t]
\vspace{-1.0cm}
\hspace{0.0cm}\psfig{figure=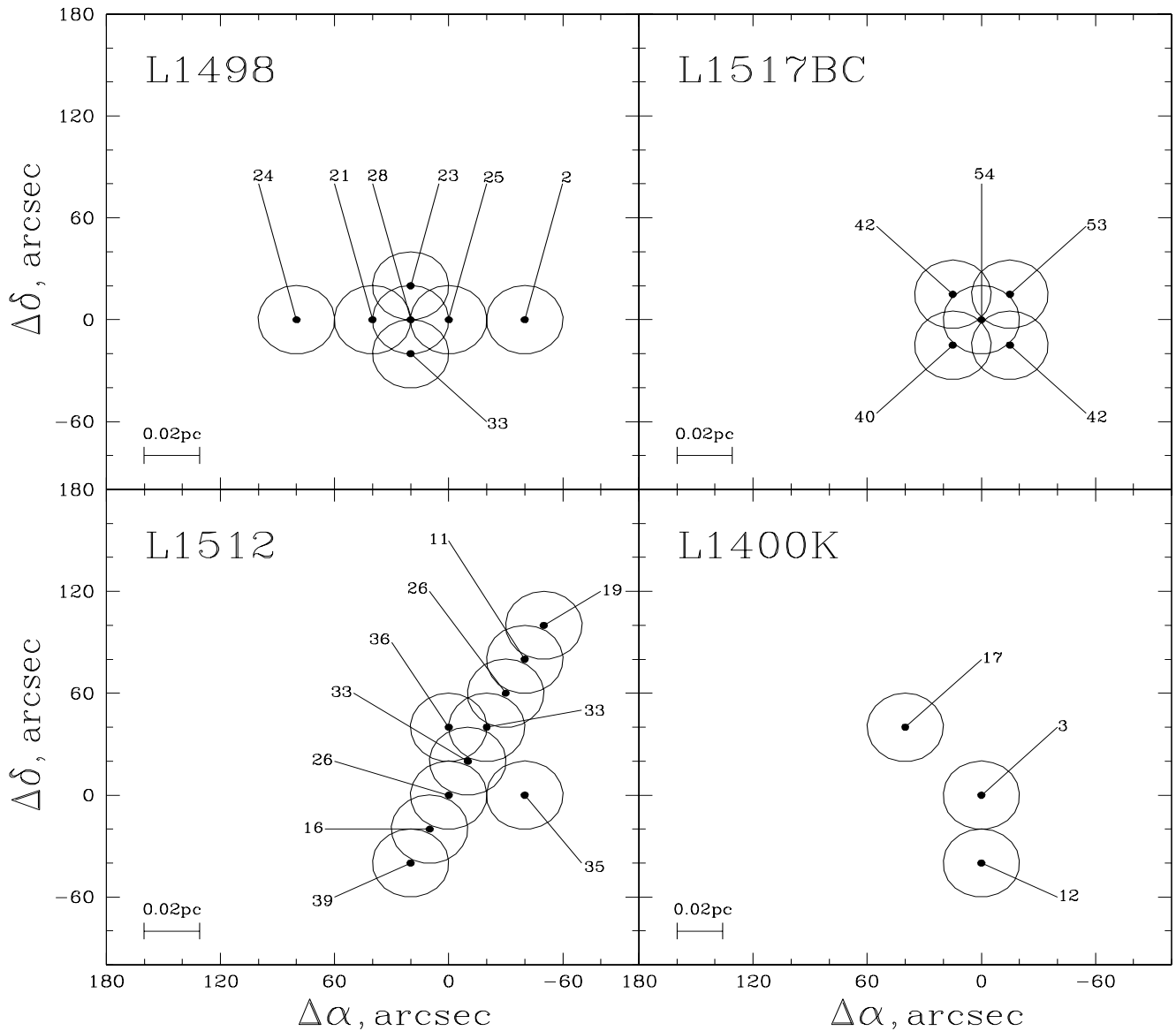,height=18cm,width=20cm}
\vspace{-1.0cm}
\caption[]{Maps of L1498, L1512, L1517BC, and L1400K showing positions 
where the radial velocity differences, \DV, between the \hcccn\ (2--1) 18.196 GHz line 
and the \nhhh\ (1,1) 23.694 GHz line have been measured 
at the Effelsberg 100-m radio telescope.
On each panel the numbers are the mean velocity shifts \DV\ (in \ms) between the
corresponding $\Delta V_s$ and $\Delta V_a$ values listed in Tables~\ref{tbl-2}--\ref{tbl-5}.
The circles show the half-power beam width at 23 GHz, whereas the linear scale is indicated
by the horizontal bars. Reference positions are given in Table~\ref{tbl-1}.
}
\label{fg-1}
\end{figure*}

\section{Observations}
\label{sect-2}

We carried out our observations targeting a sample of well-studied
nearby molecular cores that are essentially devoid of associated IR sources
(so-called starless cores with narrow emission lines).
All of our objects lie in the Taurus-Auriga-Perseus molecular complex, i.e., at the
distance of about 140 pc (L1498, L1512, L1517BC) 
and 170 pc (L1400K) from the solar system (Jijina \etal\ 1999).
Observations were performed with the 100-m Effelsberg radio telescope in January 2010.
The positions observed are listed in Table~\ref{tbl-1}. 
All sources, except L1517B, have the same central coordinates as used in Paper~I.
For L1517B, the central position is shifted at $(\Delta\alpha,\Delta\delta) = (15'',15'')$ 
with respect to the previous coordinates, and the source is called L1517BC.
The ($J,K$) = (1,1) inversion line of ammonia (NH$_3$) at 23.694 GHz and the
$J = 2-1$ rotation line of cyanoacetylene (HC$_3$N) at 18.196 GHz
were measured with a K-band HEMT (high electron mobility transistor)
dual channel receiver\footnote{At 18 GHz we used only one channel Rx2
since Rx1 was not operating during our observations.}, 
yielding spectra with angular resolution of HPBW~$\sim 40''$
in two orthogonally oriented linear polarizations. Averaging the emission from both
channels gives typical system temperatures of 100--150\,K for NH$_3$
and 80--100\,K for HC$_3$N on a main beam brightness temperature scale.

The measurements were carried out in frequency switching mode
using a frequency throw of 5\,MHz. The backend was an FFTS
operated with its minimum
bandwidth of 20\,MHz providing simultaneously 16\,384 channels
for each polarization. The resulting channel separations are 15.4 \ms\ and
20.1 \ms\ for NH$_3$ and HC$_3$N, respectively. We note,
however, that the true velocity resolution is about twice as 
lower, FWHM~$\sim 30$ \ms\ and 40 \ms, respectively (Kein \etal\ 2006).

The sky frequencies were reset at the onset of each scan and the
Doppler tracking was used continuously to track Doppler shifts during the observations.
In the Effelsberg driving program,
the position of the center of gravity of Earth is taken from the JPL
ephemerides including the influence of the Moon, and
the rotation of Earth at the telescope position
is added to this center of gravity.
The velocity of the source with respect to Local Standard of Rest (LSR),
$V_{lsr}$, is corrected for these motions and the
corrected value of $V_{lsr}$ determines the current sky frequency.
The stability of the LO-frequency during 5 minutes exposures
was about $10^{-14}$.

Observations started by measuring the continuum emission of
calibration sources (NGC\,7027, W3(OH), 3C\,286) and continued by
performing pointing measurements toward a source close to the spectroscopic
target. Spectral line measurements were interspersed with pointing
measurements at least once per hour. The calibration is estimated
to be accurate to $\pm$15\% and the pointing accuracy to be superior to
10\,arcsec.
Each molecular transition was observed in a set of four sequential 5 min exposures,
which were repeated twice at 18 GHz to compensate for the non-operating channel Rx1. 
Thus, the total exposure time at one coordinate point was one hour.
Some points were observed several times to control the accuracy of the \DV\ measurements. 
All spectra were obtained with a comparable noise level which is indicated by the
rms values in Table~\ref{tbl-1}.  
The position offsets within each individual cloud are shown in Fig.~\ref{fg-1}.

\begin{table*}[t!]
\centering
\caption{Molecular core L1498. Represented are fitting parameters 
$V_{lsr}$ (\kms)/\,b (\kms)/\,$\chi^2_\nu$. Lower limits on $T_{\rm kin}$, and
upper limits on $\sigma_{\rm turb}$ are indicated.
The numbers in parentheses are $1\sigma$ statistical errors.
The signal-to-nose ratio, S/N, per spectral channel at the maximum intensity
peak of \nhhh/\hcccn\ is indicated in the last column.
}
\label{tbl-2}
\begin{tabular}{r@{,}l c r@{.}l  r@{.}l r@{.}l r@{.}l r@{.}l c  c c}
\hline
\hline
\noalign{\smallskip}
\multicolumn{3}{c}{ } & \multicolumn{4}{c}{ NH$_3$ (1,1) } & \multicolumn{2}{c}{ HC$_3$N $(2-1)$ } & 
\multicolumn{2}{c}{ $\Delta V_s$,} & \multicolumn{2}{c}{ $\Delta V_a$,} & $T_{\rm kin}$, & $\sigma_{\rm turb}$, 
 & S/N \\[2pt]
\multicolumn{2}{c}{Offset} &  \# & \multicolumn{2}{c}{satellite} & \multicolumn{2}{c}{all} & 
\multicolumn{2}{c}{all} & \multicolumn{2}{c}{\ms} & \multicolumn{2}{c}{\ms} & \multicolumn{1}{c}{K} &
\multicolumn{1}{c}{\ms} \\[-2pt]
\multicolumn{2}{c}{${\scriptstyle (1)}$} & \multicolumn{1}{c}{${\scriptstyle (2)}$} &
\multicolumn{2}{c}{${\scriptstyle (3)}$} & \multicolumn{2}{c}{${\scriptstyle (4)}$} &
\multicolumn{2}{c}{${\scriptstyle (5)}$} & \multicolumn{2}{c}{${\scriptstyle (6)}$} &
\multicolumn{2}{c}{${\scriptstyle (7)}$} & ${\scriptstyle (8)}$ & ${\scriptstyle (9)}$ & ${\scriptstyle (10)}$\\ 
\hline 
\noalign{\smallskip}
0&0 & 1 & 7&7989(14)& 7&7994(10) & 7&8242(17) & 25&3(2.2) & 24&8(2.0) & 5.6(2.2) & 63.1(8.2) & 33\\[-2pt]
\multicolumn{2}{c}{}& & 0&110(5) & 0&116(1) & 0&099(7) &\multicolumn{2}{c}{}&\multicolumn{2}{c}{} & & &24\\[-2pt]
\multicolumn{2}{c}{}& & 1&09 & 1&07 & 0&86 & \multicolumn{2}{c}{} & \multicolumn{2}{c}{} \\
40&0 & 2 & 7&8099(10)  & 7&8089(8) & 7&8307(13)& 20&8(1.6) & 21&8(1.5) & 7.8(1.4) & 48.4(6.6) & 38\\[-2pt]
\multicolumn{2}{c}{}& & 0&110(3) & 0&111(1) & 0&085(5) &\multicolumn{2}{c}{}&\multicolumn{2}{c}{} & & &20\\[-2pt]
\multicolumn{2}{c}{}& & 1&30 & 1&13 & 1&14 & \multicolumn{2}{c}{} & \multicolumn{2}{c}{} \\
20&0 & 3 & 7&8089(9)  & 7&8089(7) & 7&8382(14)& 29&3(1.7) & 29&3(1.6) & 6.6(1.1) & 52.0(5.0) & 42\\[-2pt]
\multicolumn{2}{c}{}& & 0&113(3) & 0&109(1) & 0&087(4) &\multicolumn{2}{c}{}&\multicolumn{2}{c}{} & & & 18\\[-2pt]
\multicolumn{2}{c}{}& & 1&16 & 1&17 & 1&25 & \multicolumn{2}{c}{} & \multicolumn{2}{c}{} \\
20&0 & 8 & 7&8094(9)  & 7&8099(7) & 7&8372(17)& 27&8(1.9) & 27&3(1.8) & 5.3(2.9) & 58.1(11.9) & 43\\[-2pt]
\multicolumn{2}{c}{}& & 0&111(3) & 0&109(1) & 0&092(10) &\multicolumn{2}{c}{}&\multicolumn{2}{c}{} & & & 18\\[-2pt]
\multicolumn{2}{c}{}& & 1&13 & 1&16 & 1&17 & \multicolumn{2}{c}{} & \multicolumn{2}{c}{} \\
--40&0 & 4 & 7&7979(36)  & 7&8039(29) & 7&8032(20)& 5&3(3.3) & --0&7(3.5) & 8.3(2.2) & 60.5(6.5) & 14\\[-2pt]
\multicolumn{2}{c}{}& & 0&105(9) & 0&124(4) & 0&100(5) &\multicolumn{2}{c}{}&\multicolumn{2}{c}{} & & & 13\\[-2pt]
\multicolumn{2}{c}{}& & 1&10 & 1&11 & 0&97 & \multicolumn{2}{c}{} & \multicolumn{2}{c}{} \\
20&20 & 5 & 7&8144(10)  & 7&8149(8) & 7&8377(16)& 23&3(1.9) & 22&8(1.8) & 7.0(1.1) & 51.5(5.1) & 37\\[-2pt]
\multicolumn{2}{c}{}& & 0&112(4) & 0&110(1) & 0&087(4) &\multicolumn{2}{c}{}&\multicolumn{2}{c}{} & & & 15\\[-2pt]
\multicolumn{2}{c}{}& & 1&23 & 1&23 & 0&97 & \multicolumn{2}{c}{} & \multicolumn{2}{c}{} \\
20&--20 & 6 & 7&8019(12)  & 7&8019(9) & 7&8347(21)& 32&8(2.4) & 32&8(2.3) & 7.0(2.3) & 55.7(9.9) & 36\\[-2pt]
\multicolumn{2}{c}{}& & 0&107(3) & 0&114(1) & 0&092(8) &\multicolumn{2}{c}{}&\multicolumn{2}{c}{} & & & 13\\[-2pt]
\multicolumn{2}{c}{}& & 1&15 & 1&19 & 1&08 & \multicolumn{2}{c}{} & \multicolumn{2}{c}{} \\
80&0 & 7 & 7&7964(38)  & 7&7969(29) & 7&8207(16)& 24&3(4.1) & 23&8(3.3) & 9.3(3.2) & 46.0(14.2) & 13\\[-2pt]
\multicolumn{2}{c}{}& & 0&109(12) & 0&115(5) & 0&085(10) &\multicolumn{2}{c}{}&\multicolumn{2}{c}{} & & & 17\\[-2pt]
\multicolumn{2}{c}{}& & 0&86 & 0&85 & 0&98 & \multicolumn{2}{c}{} & \multicolumn{2}{c}{} \\
\noalign{\smallskip}
\hline
\noalign{\smallskip}
\multicolumn{16}{l}{Note.~--- Numbering in column~2 is the same as in the last column of Table~\ref{tbl-1}. }
\end{tabular}
\end{table*}

\begin{table*}[t!]
\centering
\caption{Same as Table~\ref{tbl-2} but for the molecular core L1512. 
}
\label{tbl-3}
\begin{tabular}{r@{,}l c r@{.}l  r@{.}l r@{.}l r@{.}l r@{.}l c  c c}
\hline
\hline
\noalign{\smallskip}
\multicolumn{3}{c}{ } & \multicolumn{4}{c}{ NH$_3$ (1,1) } & \multicolumn{2}{c}{ HC$_3$N $(2-1)$ } & 
\multicolumn{2}{c}{ $\Delta V_s$,} & \multicolumn{2}{c}{ $\Delta V_a$,} & $T_{\rm kin}$, & $\sigma_{\rm turb}$, 
 & S/N \\[2pt]
\multicolumn{2}{c}{Offset} &  \# & \multicolumn{2}{c}{satellite} & \multicolumn{2}{c}{all} & 
\multicolumn{2}{c}{all} & \multicolumn{2}{c}{\ms} & \multicolumn{2}{c}{\ms} & \multicolumn{1}{c}{K} &
\multicolumn{1}{c}{\ms} \\[-2pt]
\multicolumn{2}{c}{${\scriptstyle (1)}$} & \multicolumn{1}{c}{${\scriptstyle (2)}$} &
\multicolumn{2}{c}{${\scriptstyle (3)}$} & \multicolumn{2}{c}{${\scriptstyle (4)}$} &
\multicolumn{2}{c}{${\scriptstyle (5)}$} & \multicolumn{2}{c}{${\scriptstyle (6)}$} &
\multicolumn{2}{c}{${\scriptstyle (7)}$} & ${\scriptstyle (8)}$ & ${\scriptstyle (9)}$ & ${\scriptstyle (10)}$\\ 
\hline 
\noalign{\smallskip}
0&0 & 1 & 7&1015(18)& 7&1015(14) & 7&1290(21) & 27&5(2.8) & 27&5(2.5) & 10.0(2.3) & 34.6(14.8) & 24\\[-2pt]
\multicolumn{2}{c}{}& & 0&112(5) & 0&110(3) & 0&075(9) &\multicolumn{2}{c}{}&\multicolumn{2}{c}{} & & &19\\[-2pt]
\multicolumn{2}{c}{}& & 1&17 & 1&11 & 0&94 & \multicolumn{2}{c}{} & \multicolumn{2}{c}{} \\
0&0 & 10 & 7&1035(12)& 7&1045(9) & 7&1300(20) & 26&5(2.3) & 25&5(2.2) & 10.3(1.2) & 33.7(8.4) & 35\\[-2pt]
\multicolumn{2}{c}{}& & 0&108(4) & 0&111(1) & 0&075(5) &\multicolumn{2}{c}{}&\multicolumn{2}{c}{} & & &20\\[-2pt]
\multicolumn{2}{c}{}& & 1&13 & 1&11 & 1&13 & \multicolumn{2}{c}{} & \multicolumn{2}{c}{} \\
0&0 & 12 & 7&1060(12)& 7&1070(9) & 7&1320(18) & 26&0(2.2) & 25&0(2.0) & 8.9(1.8) & 42.9(9.9) & 27\\[-2pt]
\multicolumn{2}{c}{}& & 0&106(3) & 0&111(1) & 0&081(7) &\multicolumn{2}{c}{}&\multicolumn{2}{c}{} & & &16\\[-2pt]
\multicolumn{2}{c}{}& & 1&08 & 1&17 & 1&16 & \multicolumn{2}{c}{} & \multicolumn{2}{c}{} \\
0&0 & 13 & 7&1050(13)& 7&1070(12) & 7&1295(20) & 24&5(2.4) & 22&5(2.3) & 11.2(2.4) & 26.5(20.4) & 29\\[-2pt]
\multicolumn{2}{c}{}& & 0&109(4) & 0&111(3) & 0&071(10) &\multicolumn{2}{c}{}&\multicolumn{2}{c}{} & & &16\\[-2pt]
\multicolumn{2}{c}{}& & 0&95 & 0&93 & 1&13 & \multicolumn{2}{c}{} & \multicolumn{2}{c}{} \\
20&--40 & 2 & 7&1030(85)& 7&1150(52) & 7&1475(52) & 44&5(10.0) & 32&5(7.4) & 11.6(4.5) & 25.4(21.4) & 9\\[-2pt]
\multicolumn{2}{c}{}& & 0&113(12) & 0&112(12) & 0&071(8) &\multicolumn{2}{c}{}&\multicolumn{2}{c}{} & & &7\\[-2pt]
\multicolumn{2}{c}{}& & 0&94 & 0&95 & 1&07 & \multicolumn{2}{c}{} & \multicolumn{2}{c}{} \\
--20&40 & 3 & 7&0380(9)& 7&0385(8) & 7&0710(29) & 33&0(3.0) & 32&5(3.0) & 5.6(6.2) & 56.9(26.4) & 38\\[-2pt]
\multicolumn{2}{c}{}& & 0&110(3) & 0&109(1) & 0&091(22) &\multicolumn{2}{c}{}&\multicolumn{2}{c}{} & & &15\\[-2pt]
\multicolumn{2}{c}{}& & 1&07 & 1&16 & 1&06 & \multicolumn{2}{c}{} & \multicolumn{2}{c}{} \\
--40&80 & 4 & 7&0130(22)& 7&0115(17) & 7&0230(43) & 10&0(4.8) & 11&5(4.6) & 10.2(3.5) & 43.2(19.0) & 22\\[-2pt]
\multicolumn{2}{c}{}& & 0&116(5) & 0&117(3) & 0&084(13) &\multicolumn{2}{c}{}&\multicolumn{2}{c}{} & & &10\\[-2pt]
\multicolumn{2}{c}{}& & 0&92 & 0&93 & 0&70 & \multicolumn{2}{c}{} & \multicolumn{2}{c}{} \\
--30&60 & 5 & 7&0225(14)& 7&0215(10) & 7&0475(48) & 25&0(5.0) & 26&0(4.9) & 8.3(10.6) & 52.1(49.3) & 32\\[-2pt]
\multicolumn{2}{c}{}& & 0&116(4) & 0&116(1) & 0&090(38) &\multicolumn{2}{c}{}&\multicolumn{2}{c}{} & & &14\\[-2pt]
\multicolumn{2}{c}{}& & 1&09 & 1&17 & 1&02 & \multicolumn{2}{c}{} & \multicolumn{2}{c}{} \\
--10&20 & 6 & 7&0740(10)& 7&0755(8) & 7&1075(23) & 33&5(2.5) & 32&0(2.4) & 5.2(1.4) & 56.7(6.0) & 38\\[-2pt]
\multicolumn{2}{c}{}& & 0&110(5) & 0&107(1) & 0&090(5) &\multicolumn{2}{c}{}&\multicolumn{2}{c}{} & & &18\\[-2pt]
\multicolumn{2}{c}{}& & 0&91 & 1&07 & 0&87 & \multicolumn{2}{c}{} & \multicolumn{2}{c}{} \\
10&--20 & 7 & 7&1200(25)& 7&1225(18) & 7&1370(23) & 17&0(3.4) & 14&5(2.9) & 8.4(1.8) & 39.0(7.8) & 19\\[-2pt]
\multicolumn{2}{c}{}& & 0&104(7) & 0&106(4) & 0&076(5) &\multicolumn{2}{c}{}&\multicolumn{2}{c}{} & & &15\\[-2pt]
\multicolumn{2}{c}{}& & 1&09 & 1&11 & 1&09 & \multicolumn{2}{c}{} & \multicolumn{2}{c}{} \\
--40&0 & 8 & 7&0935(12)& 7&0960(9) & 7&1300(21) & 36&5(2.4) & 34&0(2.3) & 3.8(1.3) & 64.6(4.6) & 33\\[-2pt]
\multicolumn{2}{c}{}& & 0&107(3) & 0&110(1) & 0&098(4) &\multicolumn{2}{c}{}&\multicolumn{2}{c}{} & & &21\\[-2pt]
\multicolumn{2}{c}{}& & 1&01 & 1&16 & 1&05 & \multicolumn{2}{c}{} & \multicolumn{2}{c}{} \\
0&40 & 9 & 7&0505(14)& 7&0500(12) & 7&0860(23) & 35&5(2.7) & 36&0(2.6) & 8.1(1.6) & 43.4(7.2) & 27\\[-2pt]
\multicolumn{2}{c}{}& & 0&105(4) & 0&108(3) & 0&080(5) &\multicolumn{2}{c}{}&\multicolumn{2}{c}{} & & &19\\[-2pt]
\multicolumn{2}{c}{}& & 0&76 & 0&90 & 1&10 & \multicolumn{2}{c}{} & \multicolumn{2}{c}{} \\
--50&100 & 11 & 6&9875(78)& 6&9720(49) & 6&9985(53) & 11&0(9.4) & 26&5(7.2) & 18.7(4.1) & 28.9(24.3) & 11\\[-2pt]
\multicolumn{2}{c}{}& & 0&131(20) & 0&141(7) & 0&088(10) &\multicolumn{2}{c}{}&\multicolumn{2}{c}{} & & &7\\[-2pt]
\multicolumn{2}{c}{}& & 0&91 & 1&00 & 0&70 & \multicolumn{2}{c}{} & \multicolumn{2}{c}{} \\
\noalign{\smallskip}
\hline
\end{tabular}
\end{table*}

\section{Analysis}
\label{sect-3}

\begin{figure*}[t]
\vspace{0.0cm}
\hspace{-1.2cm}\psfig{figure=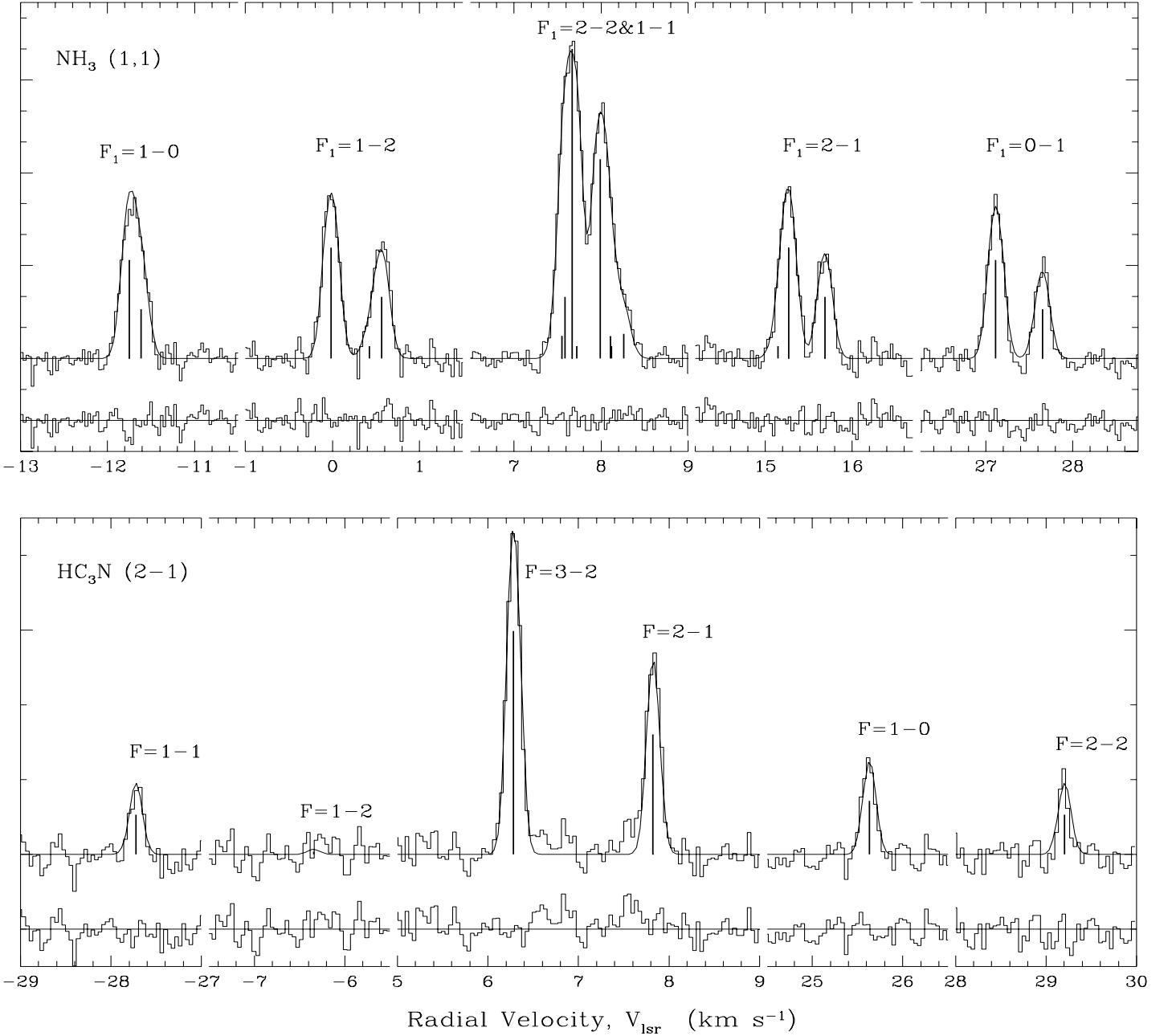,height=18cm,width=20cm}
\vspace{-0.5cm}
\caption[]{Spectra of NH$_3$ (1,1) and HC$_3$N ($2-1$) lines toward the molecular core L1498
obtained at the Effelsberg 100-m radio telescope.
The offset ($\Delta\alpha,\Delta\delta) = (0'',0'')$.
The histograms show the data averaged over two spectral channels. 
The fits are shown with solid curves.  
The vertical tick-marks indicate the positions and the relative strengths of 
the hyperfine structure components in case of local thermodynamical equilibrium
and optically thin emission.
The residual spectra are plotted below each profile. 
The fitting parameters are presented in Table~\ref{tbl-2}.
}
\label{fg-2}
\end{figure*}

We used the CLASS reduction package\footnote{http://www.iram.fr/IRAMFR/GILDAS} 
for standard data reduction. 
After corrections for the rounded frequencies (see Sec.~\ref{sect-1}),  
the individual exposures were co-added to increase 
the signal-to-noise ratio, S/N.
The spectra were folded to remove the effects
of the frequency switch, and base lines were determined for each spectrum. 
The resolved hfs components show no kinematic sub-structure and consist of an
apparently symmetric peak profile without broadened line wings or self-absorption
features, as shown in Fig.~\ref{fg-2}  where the observed
profiles of the \nhhh\ (1,1) and \hcccn\ (2--1) lines toward L1498 are plotted.
The line parameters such as the total optical depth in the transition, $\tau_{\rm tot}$ 
(i.e., the peak optical depth if all hyperfine components were placed at the same velocity),
the radial velocity, $V_{lsr}$,
the line broadening Doppler parameter, $b$, and the amplitude, $A$, were obtained through the fitting
of the one-component Gaussian model to the observed spectra as described in Paper~I:
\begin{equation}
T(v) = A\cdot \left[ 1 - \exp(-t(v)) \right]\, ,
\label{eq2}
\end{equation}
with
\begin{equation}
t(v) = \tau\cdot \sum^k_{i=1}\, a_i\, \exp\left[ -{(v - v_i - V_{lsr})^2}/{b^2} \right]\, ,
\label{eq3}
\end{equation}
which transforms for optically thin transitions into
\begin{equation}
T(v) = A'\cdot \sum^k_{i=1}\, a_i\, \exp\left[ -{(v - v_i - V_{lsr})^2}/{b^2} \right]\, .
\label{eq4}
\end{equation}
The sum in (\ref{eq3}) and (\ref{eq4}) runs over the $k = 18$ and $k = 6$ hfs components of 
the \nhhh\ $(J,K) = (1,1)$ and \hcccn\ $J = 2-1$ transitions, respectively.

The rest-frame frequencies of the \nhhh\ hfs components 
are taken from Kukolich (1967) and listed in Table~2 (Paper~I).
We checked these frequencies using the last 
JPL model\footnote{http://spec.jpl.nasa.gov/ftp/pub/catalog/catdir.html}
which takes into account the fit to all published ammonia data. 
The difference between various estimates does not exceed 0.05 kHz. 
In addition, Lapinov \etal\  (2010) carried out
new laboratory measurements of the frequencies of the \nhhh\ (1,1) hfs components 
and their results coincide with Kukolich's frequencies within the 0.2 kHz range.

The rest-frame frequencies of the \hcccn\ (2--1) hfs components listed in Table~3 (Paper~I) are
based on Lapinov (2008, private comm.) calculations which combine all available 
laboratory measurements. 
These frequencies are in agreement
with the data from the Cologne Database for Molecular Spectroscopy 
(M\"uller \etal\ 2005) 
within $1\sigma$ uncertainty interval, but Lapinov's values 
have a slightly higher precision: 
$\varepsilon_\nu \simeq 0.17$ kHz versus 0.2 to 0.7 kHz by M\"uller \etal. 
The recent laboratory measurements of the \hcccn\ (2--1) frequencies by Lapinov \etal\ (2010) did not
show any systematic shifts between the previously obtained and new measured values: 
all frequencies agree within 0.2 kHz.
Note that at the moment the accuracy with which the frequencies of \nhhh\ (1,1) and \hcccn\ (2--1) are 
known is the highest among all observable molecular transitions, with the uncertainties 
being
$\varepsilon_v$(\nhhh) $\simeq 0.6$ \ms\ ($\simeq 50$ Hz), and 
$\varepsilon_v$(\hcccn) $\simeq 2.8$ \ms\ ($\simeq 170$ Hz).

We also estimated the kinetic temperature, $T_{\rm kin}$, and the nonthermal (turbulent)
velocity dispersion, $\sigma_{\rm turb}$, based on the line broadening Doppler
parameters, $b = \sqrt{2}\sigma_v$, of the \nhhh\ (1,1) and \hcccn\ (2--1) lines. 
Here $\sigma_v$ is the line of sight velocity dispersion of the molecular gas within 
a given cloud. If the two molecular transitions trace the same material and have the same non-thermal
velocity component, then $\sigma_v$ is 
the quadrature sum of the thermal $\sigma_{\rm th}$ and turbulent $\sigma_{\rm turb}$ velocity dispersions.
In this case a lighter molecule with a mass $m_l$ should have
a wider line width as compared with a heavier molecule with $m_h > m_l$.

For thermally dominated line widths ($\beta > 1$, see Sect.~1) and co-spatially distributed species
one has the following relations (e.g., Fuller \& Myers 1993):
\begin{equation}
T_{\rm kin} = \frac{m_l m_h}{k(m_h - m_l)}(\sigma^2_l - \sigma^2_h)\, ,
\label{eq5}
\end{equation}
and 
\begin{equation}
\sigma^2_{\rm turb} = \frac{m_h \sigma^2_h - m_l \sigma^2_l}{m_h - m_l}\, ,
\label{eq6}
\end{equation}
where $k$ is Boltzmann's constant, and the thermal velocity dispersion, $\sigma_{\rm th}$, is given by
\begin{equation}
\sigma_{{\rm th},i} = (k T_{\rm kin}/m_i)^{1/2}\, .
\label{eq7}
\end{equation}
However, \hcccn\ is usually distributed in a larger volume of the molecular core as compared with \nhhh:
in general, N-bearing molecules trace the inner core, whereas C-bearing molecules occupy the outer part  
(e.g., Di Francesco \etal\ 2007).
Such a chemical differentiation and velocity gradients within the core may cause a larger nonthermal
component in the velocity distribution of \hcccn. 
If both molecules are well shielded from the external incident radiation and the gas temperature is mainly
due to the heating by cosmic rays, 
then a formal application of Eqs.(\ref{eq5}) and (\ref{eq6}) to the apparent line widths provides 
a lower limit on $T_{\rm kin}$ and an upper limit on $\sigma_{\rm turb}$.

We now consider the relative radial velocities of \hcccn\ and \nhhh, 
$\Delta V = V_{\rm rot}$(\hcccn)~-- $V_{\rm inv}$(\nhhh).
As discussed in Paper~I, the  
velocity offset \DV\ can be represented by the sum of two components
\begin{equation}
\Delta V = \Delta V_\mu + \Delta V_n,
\label{eq9}
\end{equation}
where $\Delta V_\mu$ is the shift due to a putative $\mu$-variation, and $\Delta V_n$
is the Doppler noise~--- a random component caused by possible spatial segregation
of \hcccn\ versus \nhhh\ and their different irregular Doppler shifts.
The Doppler noise can mimic or obliterate a real signal and, hence, should be
minimized. This can be achieved either by detailed mapping of the velocity field in molecular lines
which closely trace each other as performed in the present study, 
or by averaging over a large data sample as realized in Paper I.

\begin{table*}[t!]
\centering
\caption{Same as Table~\ref{tbl-2} but for the molecular core L1517BC. 
}
\label{tbl-4}
\begin{tabular}{r@{,}l c r@{.}l  r@{.}l r@{.}l r@{.}l r@{.}l c  c c}
\hline
\hline
\noalign{\smallskip}
\multicolumn{3}{c}{ } & \multicolumn{4}{c}{ NH$_3$ (1,1) } & \multicolumn{2}{c}{ HC$_3$N $(2-1)$ } & 
\multicolumn{2}{c}{ $\Delta V_s$,} & \multicolumn{2}{c}{ $\Delta V_a$,} & $T_{\rm kin}$, & $\sigma_{\rm turb}$, 
 & S/N \\[2pt]
\multicolumn{2}{c}{Offset} &  \# & \multicolumn{2}{c}{satellite} & \multicolumn{2}{c}{all} & 
\multicolumn{2}{c}{all} & \multicolumn{2}{c}{\ms} & \multicolumn{2}{c}{\ms} & \multicolumn{1}{c}{K} &
\multicolumn{1}{c}{\ms} \\[-2pt]
\multicolumn{2}{c}{${\scriptstyle (1)}$} & \multicolumn{1}{c}{${\scriptstyle (2)}$} &
\multicolumn{2}{c}{${\scriptstyle (3)}$} & \multicolumn{2}{c}{${\scriptstyle (4)}$} &
\multicolumn{2}{c}{${\scriptstyle (5)}$} & \multicolumn{2}{c}{${\scriptstyle (6)}$} &
\multicolumn{2}{c}{${\scriptstyle (7)}$} & ${\scriptstyle (8)}$ & ${\scriptstyle (9)}$ & ${\scriptstyle (10)}$\\ 
\hline 
\noalign{\smallskip}
0&0 & 1 & 5&7975(9)& 5&7990(8) & 5&8515(26) & 54&0(2.8) & 52&5(2.7) & 2.5(3.1) & 75.9(9.9) & 42\\[-2pt]
\multicolumn{2}{c}{}& & 0&115(3) & 0&118(1) & 0&111(9) &\multicolumn{2}{c}{}&\multicolumn{2}{c}{} & & &18\\[-2pt]
\multicolumn{2}{c}{}& & 1&11 & 1&09 & 0&95 & \multicolumn{2}{c}{} & \multicolumn{2}{c}{} \\
0&0 & 6 & 5&7935(9)& 5&7945(8) & 5&8490(21) & 55&5(2.3) & 54&5(2.2) & 5.9(2.1) & 62.0(8.3) & 37\\[-2pt]
\multicolumn{2}{c}{}& & 0&114(3) & 0&116(1) & 0&098(7) &\multicolumn{2}{c}{}&\multicolumn{2}{c}{} & & &23\\[-2pt]
\multicolumn{2}{c}{}& & 1&13 & 1&09 & 1&06 & \multicolumn{2}{c}{} & \multicolumn{2}{c}{} \\
15&15 & 2 & 5&7835(9)& 5&7830(8) & 5&8250(31) & 41&5(3.2) & 42&0(3.2) & 4.2(3.3) & 70.3(11.3) & 43\\[-2pt]
\multicolumn{2}{c}{}& & 0&118(3) & 0&118(1) & 0&106(10) &\multicolumn{2}{c}{}&\multicolumn{2}{c}{} & & &15\\[-2pt]
\multicolumn{2}{c}{}& & 1&21 & 1&10 & 0&71 & \multicolumn{2}{c}{} & \multicolumn{2}{c}{} \\
--15&--15 & 3 & 5&8075(9)& 5&8085(8) & 5&8500(27) & 42&5(2.8) & 41&5(2.8) & 3.4(2.6) & 70.4(9.0) & 42\\[-2pt]
\multicolumn{2}{c}{}& & 0&115(3) & 0&115(1) & 0&105(8) &\multicolumn{2}{c}{}&\multicolumn{2}{c}{} & & &18\\[-2pt]
\multicolumn{2}{c}{}& & 1&04 & 1&14 & 0&86 & \multicolumn{2}{c}{} & \multicolumn{2}{c}{} \\
15&--15 & 4 & 5&7930(9)& 5&7930(8) & 5&8325(33) & 39&5(3.4) & 39&5(3.4) & 2.3(3.0) & 73.1(9.9) & 41\\[-2pt]
\multicolumn{2}{c}{}& & 0&116(4) & 0&114(1) & 0&107(9) &\multicolumn{2}{c}{}&\multicolumn{2}{c}{} & & &13\\[-2pt]
\multicolumn{2}{c}{}& & 0&92 & 1&11 & 0&88 & \multicolumn{2}{c}{} & \multicolumn{2}{c}{} \\
--15&15 & 5 & 5&7990(13)& 5&8020(10) & 5&8535(25) & 54&5(2.8) & 51&5(2.7) & 4.9(2.8) & 64.0(10.4) & 32\\[-2pt]
\multicolumn{2}{c}{}& & 0&119(5) & 0&114(1) & 0&099(9) &\multicolumn{2}{c}{}&\multicolumn{2}{c}{} & & &18\\[-2pt]
\multicolumn{2}{c}{}& & 0&87 & 1&07 & 1&01 & \multicolumn{2}{c}{} & \multicolumn{2}{c}{} \\
\noalign{\smallskip}
\hline
\end{tabular}
\end{table*}

\section{Results}
\label{sect-4}

Equations (\ref{eq2})-(\ref{eq4}) define our model of the hfs line profiles.
The model parameters were determined by using a nonlinear least-squares method (LSM).
The best fit values of the radial velocity $V_{lsr}$ and the Doppler width $b$ 
are given in Tables~\ref{tbl-2}-\ref{tbl-5}, columns 3, 4, and 5.
An example of the fit is shown in Fig.~2. 
The residuals depicted below the profiles are normally distributed with zero mean and 
dispersion equal to the noise dispersion in the observed spectrum.
While dealing with \nhhh, we calculated two sets of the fitting
parameters: ($i$) based on the analysis of only optically thin satellite lines 
with $\Delta F_1 \ne 0$ (col.~3),
and ($ii$) obtained from the fit to the entirety of the \nhhh\ (1,1) spectrum
including the main transitions with $\Delta F_1 = 0$ 
which have optical depths $\tau \ga 1$ (col.4), 
as can be inferred from the relative intensities of the hfs components.
The corresponding radial velocity offsets are marked by $\Delta V_s$ and $\Delta V_a$.
All hfs transitions of the \hcccn\ (2--1) line from our sample are optically thin ($\tau \la 1$)  
and their $V_{lsr}$ and $b$ values are listed in col.~5.
The calculated limiting values of the kinetic temperature $T_{\rm kin}$ (lower limit)
and the nonthermal rms velocity dispersion $\sigma_{\rm turb}$ (upper limit) of the material 
presumably traced by the
\nhhh\ and \hcccn\ emission are given in columns~8 and 9.
Column~10 presents the signal-to-noise ratio per spectral 
channel at the maximum intensity peak of \nhhh/\hcccn.

The $1\sigma$ errors of the fitting parameters were
calculated from the diagonal elements of the covariance matrix 
at the minimum of $\chi^2$. In addition, the error of 
the radial velocity $V_{lsr}$ was calculated independently by the $\Delta \chi^2$ method. 
However, because of the spectral leakage (Klein \etal\ 2006), 
the fluxes in neighboring spectral channels are correlated 
(correlation coefficient $r \simeq 0.61$), and
these errors must be further corrected. 
The correlation length of $\xi = 2$ spectral channels can 
be deduced from the analysis of the autocorrelation function (ACF) of the 
intensity fluctuations in the spectral regions free from emission lines 
(computational details are given in Levshakov \etal\ 1997, 2002).
The data become less correlated ($\xi = 1$, $r \simeq 0.36$) after averaging over two neighboring spectral channels.
The LSM is known to be an unbiased and minimum variance estimator (Gauss-Markov theorem) irrespective of
the distribution of the measurements including correlated data as well (Aitken 1934).
This means that the best fit model parameters deduced from both the strongly and 
less correlated spectra should coincide. Numerical tests confirm this assumption: from both spectra we obtained
exactly the same $V_{lsr}$ and $b$ values, but in case of strongly correlated data 
the errors representing the $\Delta \chi^2 =1$ ellipsoid
were a factor of $\sim 1.3$ smaller as compared with the errors based on less correlated data.
Thus, errors given 
in parentheses in Tables~\ref{tbl-2}-\ref{tbl-5}, columns 3-5, represent calculated values
multiplied by this factor. 
These errors  were further transformed into uncertainties of the quantities listed in columns 6-9.
We also checked both folded and unfolded spectra and found that the 
measured velocity offsets are not affected by the folding procedure.

Since \hcccn\ was observed with one K-band channel (Rx2) and \nhhh\ with both (Rx1 and Rx2),
we compared the Rx2 data of \hcccn\ with \nhhh\ from the same channel and
found that changes in the radial velocities were insignificant,
within $1\sigma$ uncertainty range.

Besides the formal statistical (model fitting) uncertainties
there can be additional errors caused by instrumental imperfections.
To control this type of errors, we carried out repeated observations at 10 offsets
having the same coordinates: $\#$3, 8 (L1498), 1, 10, 12, 13 (L1512), 1, 6 (L1517BC), and
1, 2 (L1400K) in Table~\ref{tbl-1}. 
The \DV\ dispersions resulting from these repeated measurements (Tables~\ref{tbl-2}-\ref{tbl-5})
are $\sigma(\Delta V_s) = 1.6$ \ms\ and $\sigma(\Delta V_a) = 2.0$ \ms.
These dispersions are systematically lower than the reported $1\sigma$ errors of $\Delta V_s$ and $\Delta V_a$,
which ensures that we are not missing any significant instrumental errors at the level of a few \ms.

The comparison of the velocity dispersions 
determined from the \nhhh\ (1,1) and \hcccn\ (2--1) lines 
(Tables~\ref{tbl-2}-\ref{tbl-5}) does not
show any significant variations with position within each molecular core.
All data are consistent with thermally dominated line broadening, i.e., as expected,
$1 < \beta < \sqrt{3}$ for both sets  `$s$' and `$a$' of the data points.
In particular, the latter set gives the following weighted mean values:
$\beta_{\scriptscriptstyle \rm L1498} = 1.24(3)$, $\beta_{\scriptscriptstyle \rm L1512} = 1.32(5)$, 
$\beta_{\scriptscriptstyle \rm L1517BC} = 1.11(1)$, and $\beta_{\scriptscriptstyle \rm L1400K} = 1.23(4)$.
The maximum contribution of the nonthermal motions to the velocity dispersion is
observed in the L1517BC core. 
The weighted mean values of the velocity dispersions $(b/\sqrt{2})$ for \nhhh\
range between
$\sigma_{\scriptscriptstyle \rm L1512} = 78(1)$ \ms\ and
$\sigma_{\scriptscriptstyle \rm L1400K} = 86(1)$ \ms, 
and for \hcccn\ between
$\sigma_{\scriptscriptstyle \rm L1512} = 59(2)$ \ms\ and
$\sigma_{\scriptscriptstyle \rm L1517BC} = 74(1)$ \ms.
This can be compared with the speed of sound 
inside a thermally dominated region of a cold molecular core which is defined as
(e.g., Shu 1977)
\begin{equation}
v_s = (k T_{\rm kin}/m_0)^{1/2}\, ,
\label{eq8}
\end{equation}
where $m_0$ is the mean molecular mass. 
With  $m_0 \approx 2.3$ amu for molecular clouds,
one has $v_s \approx 60\sqrt{T_{\rm kin}}$ \ms, which
shows that at the typical kinetic temperature of 10 K, the nonthermal velocities 
are in general subsonic, and that the selected targets do represent the quiescent material
at different distances from the core centers.

The nonthermal velocity dispersions derived from the apparent line widths
depend on the adopted gas temperature.
The gas temperature in a molecular cloud is determined by the balance between
heating and cooling. If the only source of heating is the cosmic rays and the cooling 
is due to the line radiation mainly from  CO, then a lower bound on the kinetic
temperature is about 8 K (Goldsmith \& Langer 1978).
In cloud cores where the gas is well-shielded from the background ionizing radiation 
the gas temperature is about 10 K (Goldsmith 2001). In particular, just
this value is measured in the
L1498, L1517B and L1512 cores considered here 
(Benson \& Myers 1989; Lee \etal\ 2003; Tafalla \etal\ 2004).  

\begin{table*}[t!]
\centering
\caption{Same as Table~\ref{tbl-2} but for the molecular core L1400K. 
}
\label{tbl-5}
\begin{tabular}{r@{,}l c r@{.}l  r@{.}l r@{.}l r@{.}l r@{.}l c  c c}
\hline
\hline
\noalign{\smallskip}
\multicolumn{3}{c}{ } & \multicolumn{4}{c}{ NH$_3$ (1,1) } & \multicolumn{2}{c}{ HC$_3$N $(2-1)$ } & 
\multicolumn{2}{c}{ $\Delta V_s$,} & \multicolumn{2}{c}{ $\Delta V_a$,} & $T_{\rm kin}$, & $\sigma_{\rm turb}$, 
 & S/N \\[2pt]
\multicolumn{2}{c}{Offset} &  \# & \multicolumn{2}{c}{satellite} & \multicolumn{2}{c}{all} & 
\multicolumn{2}{c}{all} & \multicolumn{2}{c}{\ms} & \multicolumn{2}{c}{\ms} & \multicolumn{1}{c}{K} &
\multicolumn{1}{c}{\ms} \\[-2pt]
\multicolumn{2}{c}{${\scriptstyle (1)}$} & \multicolumn{1}{c}{${\scriptstyle (2)}$} &
\multicolumn{2}{c}{${\scriptstyle (3)}$} & \multicolumn{2}{c}{${\scriptstyle (4)}$} &
\multicolumn{2}{c}{${\scriptstyle (5)}$} & \multicolumn{2}{c}{${\scriptstyle (6)}$} &
\multicolumn{2}{c}{${\scriptstyle (7)}$} & ${\scriptstyle (8)}$ & ${\scriptstyle (9)}$ & ${\scriptstyle (10)}$\\ 
\hline 
\noalign{\smallskip}
0&0 & 1 & 3&2535(27)& 3&2540(20) & 3&2610(40) & 7&5(4.8) & 7&0(4.5) & 6.3(5.3) & 66.3(18.9) & 20\\[-2pt]
\multicolumn{2}{c}{}& & 0&121(5) & 0&122(3) & 0&104(16) &\multicolumn{2}{c}{}&\multicolumn{2}{c}{} & & &12\\[-2pt]
\multicolumn{2}{c}{}& & 1&15 & 1&17 & 0&98 & \multicolumn{2}{c}{} & \multicolumn{2}{c}{} \\
0&0 & 2 & 3&2485(25)& 3&2475(20) & 3&2465(34) & --2&0(4.2) & --1&0(3.9) & 5.5(2.5) & 66.5(8.2) & 19\\[-2pt]
\multicolumn{2}{c}{}& & 0&112(7) & 0&119(3) & 0&103(7) &\multicolumn{2}{c}{}&\multicolumn{2}{c}{} & & &13\\[-2pt]
\multicolumn{2}{c}{}& & 0&86 & 0&85 & 0&98 & \multicolumn{2}{c}{} & \multicolumn{2}{c}{} \\
0&--40 & 3 & 3&2135(26)& 3&2160(21) & 3&2270(26) & 13&5(3.7) & 11&0(3.3) & 9.0(1.9) & 58.7(6.5) & 18\\[-2pt]
\multicolumn{2}{c}{}& & 0&132(5) & 0&125(3) & 0&099(5) &\multicolumn{2}{c}{}&\multicolumn{2}{c}{} & & &17\\[-2pt]
\multicolumn{2}{c}{}& & 1&16 & 1&11 & 0&98 & \multicolumn{2}{c}{} & \multicolumn{2}{c}{} \\
40&40 & 4 & 3&3540(27)& 3&3570(20) & 3&3720(31) & 18&0(4.1) & 15&0(3.7) & 12.6(4.1) & 32.9(29.0) & 22\\[-2pt]
\multicolumn{2}{c}{}& & 0&126(7) & 0&120(3) & 0&079(16) &\multicolumn{2}{c}{}&\multicolumn{2}{c}{} & & &15\\[-2pt]
\multicolumn{2}{c}{}& & 1&01 & 1&10 & 1&08 & \multicolumn{2}{c}{} & \multicolumn{2}{c}{} \\
\noalign{\smallskip}
\hline
\end{tabular}
\end{table*}

In the L1498 cloud, we obtain a lower limit on the kinetic temperature 
$T_{\rm kin} = 7.1\pm0.5$ K (average over 8 points) which is slightly lower
than $T_{\rm kin} = 10$ K measured by a different method from the 
relative population of the $(J,K) = (2,2)$ and (1,1) levels of \nhhh\ 
described by the rotational temperature $T^{21}_{\scriptscriptstyle \rm R}$
(Tafalla \etal\ 2004). 
For L1512, the temperature averaged over 11 points (Table~\ref{tbl-3})
is $T_{\rm kin} = 9.6\pm0.6$ K which is consistent with the value of 10 K. 
The kinetic temperatures in the L1517BC core
is well below the Tafalla et al.'s value of 9.5 K 
in all scanned points which means that the non-thermal
velocity dispersions of \nhhh\ and \hcccn\ differ significantly and that both
species do not trace the same material. The three scanned points in the L1400K core
show $T_{\rm kin} = 8.3\pm3.2$ K~-- close to the expected value (measurements
of the gas temperature in this core were not performed in previous studies). 
We note, however, that when the linewidths of \nhhh\ and \hcccn\ are
comparable, the estimate by means of Eq.~(\ref{eq5})  becomes unstable and leads to
unphysical results (e.g., points $\#6$, 8 in Table~\ref{tbl-3}).
In general, the spatial fluctuations of $T_{\rm kin}$ do
not exceed a few kelvin implying the uniform heating and
the absence of the localized heat sources. 
The kinetic temperature tends to rise with the distance
from the core center (points $\#7$ in Table~\ref{tbl-2},
$\#11$ in Table~\ref{tbl-3}, and $\#4$ in Table~\ref{tbl-4}) which is in line with
the results of Tafalla \etal\ (2004). 

\begin{figure*}
\vspace{0.0cm}
\hspace{-1.2cm}\psfig{figure=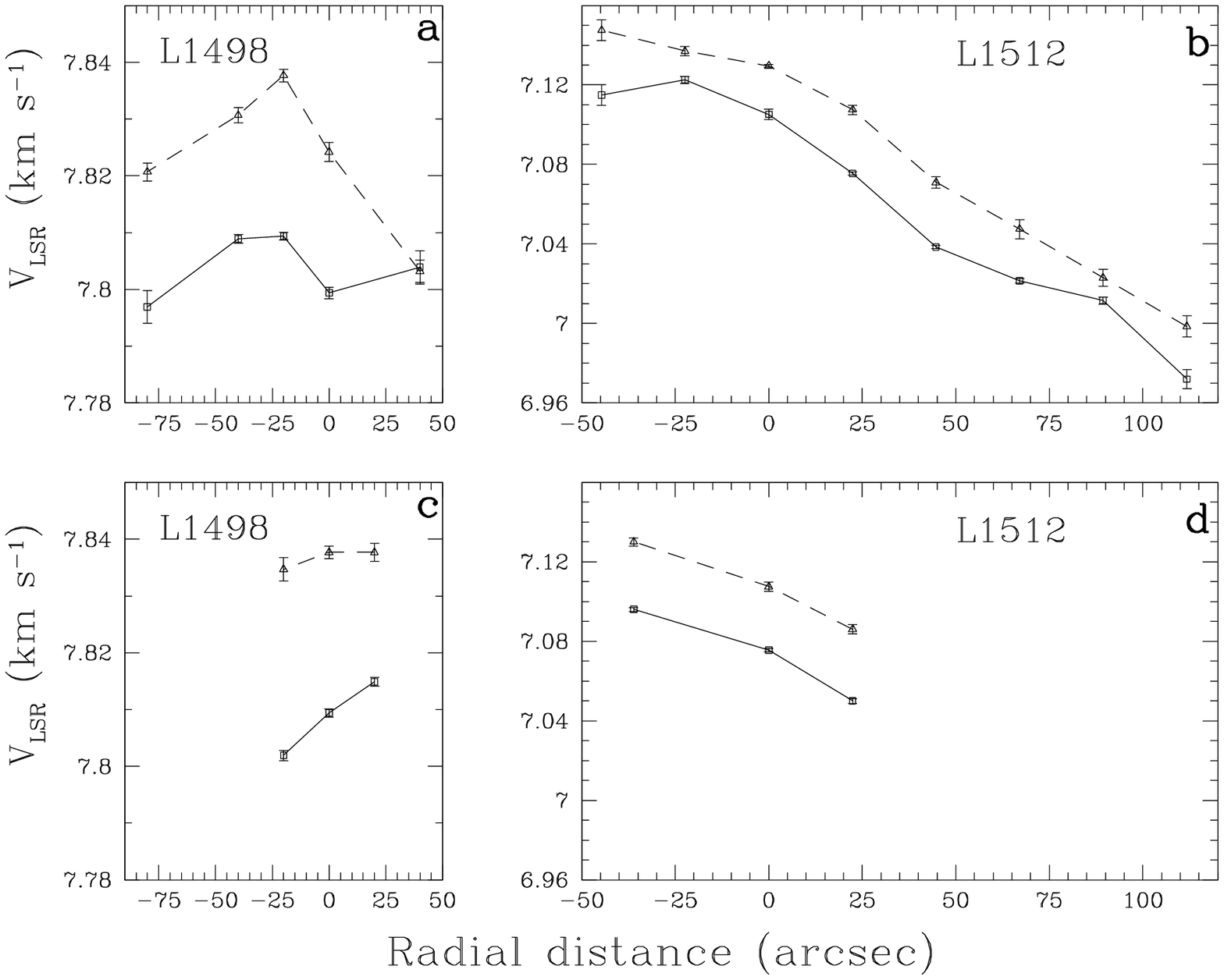,height=15cm,width=20cm}
\vspace{-3.0cm}
\caption[]{The line of sight velocities of
\nhhh\ $(J,K) = (1,1)$ (squares) and \hcccn\ $J = 2-1$ (triangles)
at different radial distances along the main diagonal cuts (panels {\bf a} and {\bf b})
and in the perpendicular directions (panels {\bf c} and {\bf d})
of the molecular cores L1498 and L1512 shown in Fig.~\ref{fg-1}. 
The reference point in panels {\bf a} and {\bf b} is 
$(\Delta\alpha,\Delta\delta) = (0'',0'')$, but it is
$(20'',0'')$ and $(-10'',20'')$ in panels {\bf c} and {\bf d}, respectively.
The radial distances of the points from the circular sector with the central angle $0^o \leq \psi < 180^o$ 
are positive, whereas those from the sector $180^o \leq \psi < 360^o$ are negative.
For multiple observations at the same coordinate the mean values of $V_{lsr}$  are depicted.
The error bars show the $1\sigma$ uncertainties. 
}
\label{fg-3}
\end{figure*}

\begin{figure*}
\vspace{0.0cm}
\hspace{-1.2cm}\psfig{figure=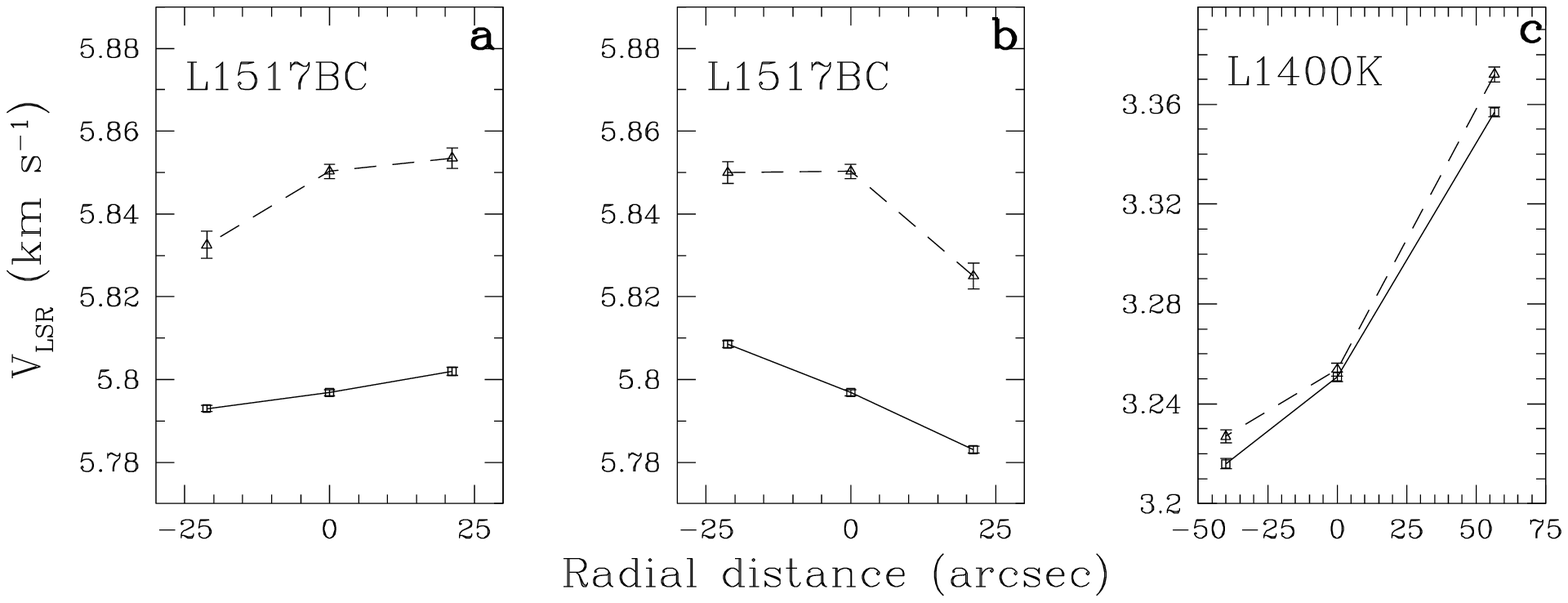,height=13cm,width=20cm}
\vspace{-7.0cm}
\caption[]{The line of sight velocities of
\nhhh\ $(J,K) = (1,1)$ (squares) and \hcccn\ $J = 2-1$ (triangles)
at different radial distances along the perpendicular diagonal cuts 
of the molecular core L1517BC (panels {\bf a} and {\bf b}),
and along three points of L1400K (panel {\bf c}) shown in Fig.~\ref{fg-1}. 
The origin of coordinates is
$(\Delta\alpha,\Delta\delta) = (0'',0'')$ in each panel.
The radial distances of the points from the circular sector with the central angle $0^o \leq \psi < 180^o$ 
are positive, whereas those from the sector $180^o \leq \psi < 360^o$ are negative.
For multiple observations at the same coordinate the mean values of $V_{lsr}$  are depicted.
The error bars show the $1\sigma$ uncertainties. 
}
\label{fg-4}
\end{figure*}

The radial velocity profiles along the different diagonal cuts toward the selected 
targets are depicted in Figs.~\ref{fg-3} and \ref{fg-4}.
The diagonal cut in L1498
between $r= -80''$ and $r = 40''$ (Fig.~\ref{fg-3}{\bf a})
exhibits coherently changing velocities of $V_{lsr}$(\hcccn) and $V_{lsr}$(\nhhh)
(except the point $r = 40''$ at the core edge) with a turn at $r = 20''$. 
The velocity gradient is small,
$|\nabla V_{lsr}| \approx 0.5$ km~s$^{-1}$~pc$^{-1}$. 
This picture coincides with the previously obtained results based on observations of
CO, CS, N$_2$H$^+$, and \nhhh\ in this core 
and was interpreted as an inward flow
(Lee \etal\ 2001; Tafalla \etal\ 2004).
Taking together, all available observation classify the L1498 cloud
as one of the most quiet molecular core. Thus, we can expect that in this core the Doppler noise,
i.e., the irregular random shifts in the radial velocities between different transitions, 
is minimal.

In the L1512 core, the $V_{lsr}$(\nhhh) and $V_{lsr}$(\hcccn) distributions are almost
parallel (Fig.~\ref{fg-3}{\bf b}, {\bf d}). 
The same kinematic picture was obtained for this core by Lee \etal\ (2001) from observations of
CS and \nhhh\ lines and interpreted as a simple rotation around the center.
The velocity gradients derived from both the \nhhh\ and \hcccn\ lines are similar,
$\nabla V_{lsr} \approx 1.5$ km~s$^{-1}$~pc$^{-1}$, and  consistent with the
gradient based on N$_2$H$^+$ measurements by Caselli \etal\ (2002). This
means that \nhhh, \hcccn, and \nnhp\ trace the same gas and the Doppler shifts \DV\ 
between them should be insignificant.

In the core L1517BC which is known to be very compact (Lee \etal\ 2001) we observed only a small
central part $30''\times30''$ where the velocities of \nhhh\ and \hcccn\ along two perpendicular cuts
(Fig.~\ref{fg-4}{\bf a}, {\bf b}) do not change much:
$\nabla V_{lsr}$(\nhhh) $\approx 0.3$ km~s$^{-1}$~pc$^{-1}$, 
$\nabla V_{lsr}$(\hcccn) $\approx 0.8$ km~s$^{-1}$~pc$^{-1}$ (panel {\bf a}),  
and
$\nabla V_{lsr}$(\nhhh) $\approx \nabla V_{lsr}$(\hcccn) $\approx -0.8$ km~s$^{-1}$~pc$^{-1}$
(panel {\bf b}).
However, a wider area ($\approx 80''\times80''$)
observation of this core revealed 
an outward gas motion at the core periphery with 
a higher velocity gradient,
$\nabla V_{lsr}$(N$_2$H$^+$) $\approx 1.1$ km~s$^{-1}$~pc$^{-1}$ (Tafalla \etal\ 2004)
which is consistent with earlier results on \nhhh\ observations by Goodman \etal\ (1993).
This can lead to an additional shift in the radial velocity of \hcccn\ line since, 
in general, the C-bearing molecules occupy a lower density gas ($n \sim 10^3$ \cmm) in the envelope
of the molecular core. A higher non-thermal velocity dispersion of the \hcccn\ line as compared to \nhhh\ 
was already mentioned above in regard with the temperature measurements in this core.

In the L1400K core we observed only three positions. Both molecules trace the same gradient of
$\nabla V_{lsr} \approx 1.9$ km~s$^{-1}$~pc$^{-1}$ which is in line with
$\nabla V_{lsr}$(N$_2$H$^+$) $= 1.8\pm0.1$ km~s$^{-1}$~pc$^{-1}$ derived by Caselli \etal\ (2002).
The mapping in different molecular lines by Tafalla \etal\ (2002) revealed that
L1400K deviates significantly from spherical symmetry and exhibits quite complex
kinematic structure. In particular, distributions of \nnhp\ and \nhhh\ do not coincide:
\nnhp\ has an additional component to the west from the center.
This explains why
Craspi \etal\ (2005) report for this core a blue-ward skewness,
$\theta = -0.42\pm0.10$, of the the N$_2$H$^+$ (1--0) hfs profiles,
whereas in our observations the \nhhh\ (1,1) hfs transitions are fully
symmetric: 
at the central position,
the skewness of the resolved and single hfs component $F_1 F = 0\,\frac{1}{2} \rightarrow 1\,1\frac{1}{2}$ 
of \nhhh\ is $\theta = -0.1\pm0.3$.

The mean values of the velocity offsets \DV\ between \hcccn\ and \nhhh\
measured across each of 4 cores are presented in Table~\ref{tbl-6}.
For each individual molecular core we calculated the unweighted and weighted
mean \DV\ (weights inverse proportional to the variances)
by averaging over available offsets for two datasets: based on only
optically thin hfs transitions of \nhhh\ (marked by index $s$), and on the all transitions
including the strong main components of \nhhh\ (marked by index $a$).
We also used a robust redescending $M$-estimate (Maximum-likelihood) for the mean
and the normalized Median Absolute Deviation ($1.483\cdot$MAD) for the scale of
the distribution as described in the Appendix of Paper~I.
These statistics work well for inhomogeneous data sets with outliers and deviations from normality.
The median values, which are robust estimates independent on the assumed distribution of data,
are also presented to demonstrate the consistency of different calculations of the mean \DV\ values.

We obtain very similar velocity shifts 
$\langle \Delta V_a \rangle = 25.8 \pm 1.7$ \ms\ and $28.0 \pm 1.8$ \ms\ ($M$-estimates) 
for, respectively, the cores
L1498 and L1512 where the minimal level of the Doppler noise is expected. 
A larger shift $\langle \Delta V_a \rangle = 46.9\pm3.3$ \ms\ is observed in the
L1517BC core~--
again in accord with the revealed kinematic
structure of this core which allows us to expect a higher radial velocity for the \hcccn\ line.
On the other hand, a lower value $\langle \Delta V_a \rangle = 8.5\pm3.4$ \ms\ in L1400K may be
due to irregular kinematic structure of
the core center which could increase the radial velocity of the \nhhh\ line.
Thus, as the reference velocity offset we choose the most robust $M$-estimate of the mean value from 
the L1498 and L1512 cores: 
$\langle \Delta V_a \rangle = 26.9\pm1.2_{\rm stat}$ \ms.
Taking into account that the uncertainty of the \hcccn\ (2--1)
rest frequency is about 3 \ms, whereas that of \nhhh\ (1,1) is less than 1 \ms,
we finally have
$\langle \Delta V_a \rangle = 26.9\pm1.2_{\rm stat}\pm3.0_{\rm sys}$ \ms.
Being interpreted in terms of the electron-to-proton
mass ratio variation, as defined in Eq.(\ref{eq1}),
this velocity offset provides \dmm\ = $26\pm1_{\rm stat}\pm3_{\rm sys}$ ppb. 

\begin{figure}
\vspace{-0.5cm}
\hspace{-5.5cm}\psfig{figure=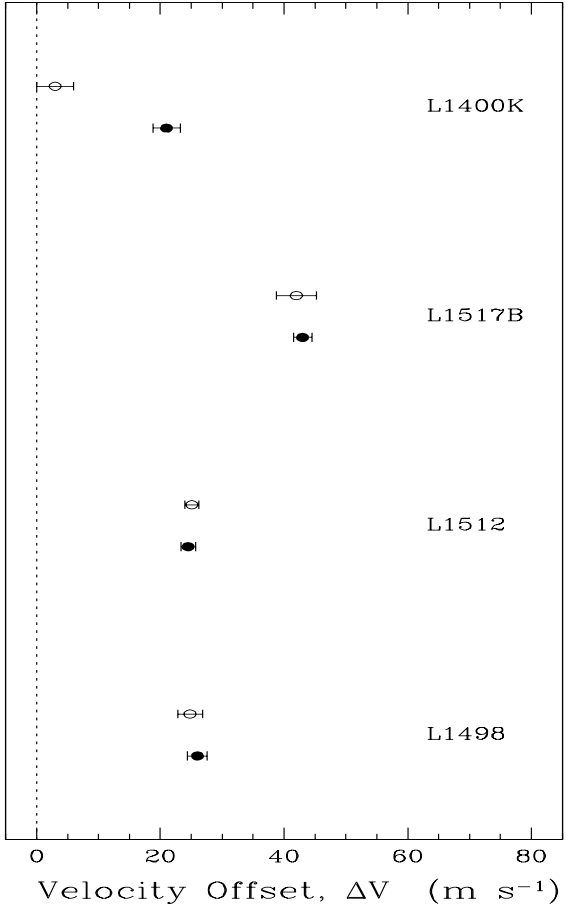,height=14cm,width=20cm}
\vspace{-0.5cm}
\caption[]{Comparison of the relative radial velocities between 
the \hcccn\ (2--1) and \nhhh\ (1,1) lines measured at the same 
coordinates in Feb 2009 (filled circles) and in Jan 2010 (open circles).
The 2009 data are corrected by 4 \ms\ as described in Sect.~\ref{sect-1}.
For L1517B the offset ($\Delta \alpha, \Delta \delta)_{2010}$ = $(15'',15'')$
corresponds to the offset ($\Delta \alpha, \Delta \delta)_{2009}$ = $(0'',0'')$.
For multiple observations at the same coordinate the mean value is depicted.
The error bars show the $1\sigma$ uncertainties. 
}
\label{fg-5}
\end{figure}

In Fig.~\ref{fg-5}, we compare the velocity offsets \DV\  obtained in our observations 
with the 100-m Effelsberg telescope in Feb 2009 and Jan 2010. 
For L1498, L1512 and L1517BC, the reproducibility is very good, whereas
for L1400K the results differ significantly. The reason for this discrepancy is unclear.

\section{Discussion}
\label{sect-5}

\subsection {Kinematic velocity shifts}
\label{sect-5-1}

In two molecular cores, L1498 and L1512, with lowest Doppler noise, we
register very close values of the velocity offset \DV\ $\sim 27$ \ms\ between the
rotational transition \hcccn\ (2--1) and inverse transition \nhhh\ (1,1),
and these values coincide with the most accurate estimate obtained from 
the Effelsberg dataset on 12 molecular clouds in the Milky Way (Paper~I).
Two other cores, L1517B and L1400K, exhibit velocity
shifts which are either higher ($\sim47$ \ms\ in L1517B) or lower ($\sim 9$ \ms\ in L1400K) 
than the mean value, but the positive (L1517B) and
negative (L1400K) deflections from the mean can be explained from the observed kinematics
in these cores. 

The rotational transition \hcccn\ (2--1) was chosen for several reasons: its laboratory
frequency is known with an accuracy of $\sim 3$ \ms, 
it closely traces the \nhhh\ (1,1) emission,
and its frequency of 18.2 GHz is close to 23.7 GHz of \nhhh\ (1,1) yielding the similar angular resolution
for both species. However, it is well known that the C-bearing molecules freeze out in the central
core parts whereas the N-bearing molecules survive. 
Thus, observing the \nhhh\ and \hcccn\ molecules we cannot completely
reject the possibility that the velocity shifts between them are of kinematic nature.

It is clear that the simultaneous observation of the 
\nhhh\ (1,1) inverse transition  and rotational
transitions of some N-bearing molecules such as, e.g., \nnhp\ (1-0) and \nndp\ (1-0) would
give a more stringent test. The main obstacle on this way is that the laboratory frequencies of
\nnhp\ and \nndp\ are known with accuracy not better than 14 \ms\ (Paper~I, Sect.4.2). Using
the \nnhp\ rest frequency from the Cologne Database for Molecular Spectroscopy 
(CDMS; M\"uller \etal\ 2005) 
and observing with the Nobeyama 45-m telescope we obtained a
velocity shift between \nnhp\ and \nhhh\  of $23.0\pm3.4$ \ms\ in L1498,  $24.5\pm4.3$ \ms\ in L1512,
and $21.0\pm5.1$ \ms\ in L1517B (Paper I). 
We note that similar shifts are indicated for the central parts of L1498
and L1517B in Fig.~11 from Tafalla \etal\ (2004) who observed these targets with the 30-m IRAM telescope and
used the \nnhp\ rest frequency very close to the CDMS value.

In the prestellar core L183, Pagani \etal\ (2009) observed
\nnhp\ (1--0) and \nndp\ (1--0) and found that along two perpendicular cuts the two lines
are systematically shifted with respect to \nhhh\ (1,1).
The velocity offset is about 28 \ms,
when adopting the CDMS catalogue rest frequencies
(Paper~I; Molaro \etal\ 2009), but $\Delta V \approx 40$ \ms\ if the frequency of \nnhp\
is taken from Caselli \etal\ (1995).

Thus, almost the same velocity offset between the \nhhh\ inversion transition 
and rotational lines from different molecular species (\hcccn,  \nnhp, \nndp),
different targets, and different telescopes is observed.
Tentatively, this supports our initial hypothesis that 
the equivalence principle of local position invariance
may break in low-density environment.
Obviously, for more definite conclusions new laboratory measurements of the
rest frequencies and new observations involving
other targets and other rotational transitions are required.

In this context we note that the matter density in diffuse interstellar clouds,
where the UV lines of the molecular hydrogen H$_2$ are observed, is about
5 orders of magnitude lower than in molecular cores. This implies that the
expected value of \dmm\ based on the H$_2$ line position measurements should
not be less than $3\times10^{-8}$.
This makes future experiments with CODEX and ESPRESSO spectrograph for extremely stable
Doppler measurements (Pasquini \etal\ 2009) of a great importance
since current limits on \dmm\ obtained with the UVES/VLT and HERES/Keck are
at the level of a few ppm (e.g., Malec \etal\ 2010).

\subsection{Frequency shifts caused by external fields}
\label{sect-5-2}

In this section we consider possible systematic effects caused by external fields:
we start with a static electric field, 
then consider a thermal background radiation, and, finally, discuss a static 
magnetic field. We deduce all estimates for the inversion transition in ammonia. 
Rotational transitions can be treated in a similar way.

\subsubsection{Stark effect induced by static electric fields}
\label{sect-5-2-1}

Static electromagnetic fields are always present and can affect
the observed frequencies of microwave transitions in the laboratory and
in astrophysical environments (see, e.g., Bethlem \etal\ 2008). 

The static Stark displacement of the inversion frequency in \nhhh\ is given by 
(Townes \& Schawlow 1955)
\begin{equation}
\delta \omega_\mathrm{inv}
  =2\frac{\langle {\cal D}\rangle^2 {\cal F}^2}{\hbar^2 \omega_\mathrm{inv}}
  =2\left[\frac{M K}{J(J+1)}\right]^2
 \frac{{\cal D}^2 {\cal F}^2}{\hbar^2 \omega_\mathrm{inv}}\,.
\label{stark1}
\end{equation}
Here $\hbar = h/2\pi$ is the reduced Planck constant, 
$\langle {\cal D}\rangle$ is the electric dipole amplitude for the \nhhh\ (1,1) inversion 
transition, ${\cal D} = 1.42$~D is the molecular dipole moment, 
${\cal F}$ is the electric field strength, 
$J$, $M$, and $K$ are, respectively, 
the molecular angular momentum and its projections on the 
external electric field and on the molecular axis.
Averaging over the projection of $M$, one finds\footnote{The projection of $M$ ranges between  $-J$ and $J$.
Assuming equal population of these levels (all of them  have the same energy), we have 
$\langle M^2 \rangle = \frac{1}{2J+1}\ \sum_{M=-J}^{J}\ M^2 = J(J+1)/3.$ }:
\begin{equation}
  \frac{\delta V_\mathrm{St}}{c}
  =\frac{\delta \omega_\mathrm{inv}}{\omega_\mathrm{inv}}
  =\frac{2K^2}{3J(J+1)}
   \left(\frac{\cal D F}{\hbar \omega_\mathrm{inv}}\right)^2,
\label{stark2}
\end{equation}
where $\delta V_\mathrm{St}$ is the line-of-sight velocity offset caused by the Stark effect. 
The largest offset occurs for the $K=J$ transitions. In this case the shift of
$\delta V_\mathrm{St} = 1$ \ms\ requires a field of ${\cal F} \approx 2$~V~cm$^{-1}$.
For laboratory conditions, this is a sufficiently large field which is 
easily controlled. For dense interstellar molecular clouds, 
such a field would accelerate electrons 
to ultra-relativistic energies ($\sim 10^{10}$ eV) within their mean free path
($\ell_{\rm e} \sim 5\times10^{14}/n_{{\scriptscriptstyle \rm H}_2}$ cm).
Thus, we conclude 
that static electric fields cannot cause any systematic effects on this scale.

We note that the collision shift when molecules get close to each other and the
\nhhh\ (1,1) inversion line is shifted due to the electric dipole-induced 
interactions between collision partners is also less than 1 \ms\ for the physical
conditions realized in the most accurate laboratory studies with molecular
fountains and in cold interstellar molecular cores (Bethlem \etal\ 2008).

\begin{table*}[t!]
\centering
\caption{Mean velocity offsets \DV\ (in \ms) between the \hcccn\ (2--1)
and \nhhh\ (1,1) lines. The values in parentheses are the $1\sigma$ uncertainties. 
}
\label{tbl-6}
\begin{tabular}{lcccccccccc}
\hline
\hline
\noalign{\smallskip}
& \multicolumn{2}{c}{L1498} & \multicolumn{2}{c}{L1512} & \multicolumn{2}{c}{ L1517BC } & 
\multicolumn{2}{c}{ L1400K} & \multicolumn{2}{c}{ Average } \\
& $\langle\Delta V_s\rangle$ &$\langle\Delta V_a\rangle$ 
& $\langle\Delta V_s\rangle$ &$\langle\Delta V_a\rangle$ & $\langle\Delta V_s\rangle$ &$\langle\Delta V_a\rangle$
& $\langle\Delta V_s\rangle$ &$\langle\Delta V_a\rangle$ & $\langle\Delta V_s\rangle$ &$\langle\Delta V_a\rangle$ \\
\hline 
\noalign{\smallskip}
Unweighted &23.6(2.9)&22.7(3.6)&27.0(2.8)&26.6(2.0)&47.9(3.1)&46.9(2.7)&9.3(4.3)&8.0(3.4)&27.9(2.6)&27.1(2.5)\\
Weighted &24.8(2.2)&24.8(2.4)&28.5(1.9)&27.2(1.8)&49.2(3.0)&48.4(2.7)&9.8(4.3)&8.5(3.4)&29.6(2.2)&28.6(2.1)\\
$M$-estimate &26.0(1.6)&25.8(1.7)&27.8(2.3)&28.0(1.8)&47.9(3.9)&46.9(3.3)&9.8(4.5)&8.5(3.4)&27.5(2.4)& 27.3(1.5) \\
Median &24.8&24.3&26.5&26.5&48.3&46.8&10.5&9.0&26.5&26.5\\
Sample size & 8 & 8 & 13 & 13 & 6 & 6 & 4 & 4 & 31 & 31\\
\noalign{\smallskip}
\hline
\noalign{\smallskip}
\end{tabular}
\end{table*}

\subsubsection{Stark effect induced by black body radiation}
\label{sect-5-2-2}

Now we estimate the frequency shift due to the black body radiation (BBR) at a 
given radiation temperature $T$. 
According to Farley \& Wing (1981), the BBR-induced Stark shift is given by:
\begin{equation}
  \delta \omega_\mathrm{inv}
  =\frac{4\left(kT\right)^3}{\pi c^3 \hbar^4}
  \langle {\cal D}\rangle^2
   {\phi}\left(\frac{\hbar \omega_\mathrm{inv}}{kT}\right)\,,
\label{stark3}
\end{equation}
where $k$ is the Boltzmann constant, and $c$ is the speed of light. 
In Eq.(\ref{stark3}), we take into account that the levels of the inversion doublet are
shifted from each other and sum over Cartesian components of the BBR field.
The universal function ${\phi}$ has the form
\begin{equation}
 {\phi }(y) = \int_0^\infty
 \left(\frac{1}{y+x}+\frac{1}{y-x}
 \right)\frac{x^3\mathrm{d} x}{\mathrm{e}^x-1}\,,
\label{stark4}
\end{equation}
and is restricted to the interval $-2 < {\phi} <3$. 
At the radiation temperature $T \ga 3$~K, we are interested in the limit 
$y\equiv (\hbar \omega_\mathrm{inv}/kT) \ll 1$, where
\begin{equation}
 {\phi}(y)|_{y\ll 1} \approx - \frac{\pi^2 y}{3}\,.
\label{stark5}
 \end{equation}
Using this expression and averaging again over the projection of $M$, we obtain:
\begin{equation}
 \frac{\delta \omega_\mathrm{inv}}{\omega_\mathrm{inv}}
 =-\frac{4\pi }{9 c^3 \hbar^3}\frac{K^2 {\cal D}^2}{J(J+1)}
 \left(kT\right)^2\,.
\label{stark6}
\end{equation}

We see, that the BBR shift grows with temperature and it
is larger for a warm laboratory environment as compared with the interstellar 3~K
background field. 
However, even for $T\sim 300$~K the BBR shift for the inversion transition is negligible,
$\delta\omega_\mathrm{inv}/\omega_\mathrm{inv}\sim 1.5\times 10^{-13}$.
The BBR shift for the $J = 2-1$  rotational transition of the HC$_3$N molecule 
is of the same order of magnitude.

\subsubsection{Zeeman effect induced by static magnetic fields}
\label{sect-5-2-3}

Finally, we estimate the effect of a static magnetic field.
Dense molecular cores have weak magnetic fields
with typical strengths of $B \sim 10$ $\mu$G or less (Crutcher \etal\ 2010).
For example, CCS Zeeman observations of L1498 set an upper limit on
$B = 48\pm31$ $\mu$G for the line of sight component of the magnetic field 
(Levin \etal\ 2001).
The line shapes can be sensitive to random motions at the Alfv\'en velocity
associated with the magnetic field $B$,
$v_{\scriptscriptstyle \rm A} = B/\sqrt{4\pi\rho}$
(here $B$ is the strength in G of the line-of-sight component of the
magnetic vector, $\rho$ is the ion density in g~\cmm,
and $v_{\scriptscriptstyle \rm A}$ is in \cms).
The presence of the magnetic field can be felt by the neutral material through collisions
with charged particles.
If we assume that in our observations the nonthermal line widths
are due to Alfv\'enic turbulent motions whose kinetic energy is equal to
the static magnetic field energy, so that
$v_{\scriptscriptstyle \rm A} = \sqrt{3}\, v_{\rm turb}$,
then an order of magnitude estimate of the magnetic field strength at
$n_{{\scriptscriptstyle \rm H}_2} \sim 10^5$ \cmm, and
$v_{\rm turb} < 90$ \ms\ (see Sect.~\ref{sect-3})
provides an upper limit on $B < 20$ $\mu$G.
Since in this estimate we substitute the ion density by the total gas density,
the strength of the static magnetic field in the interiors of L1498, L1512, L1517B, and L1400K
should be considerably lower than this upper limit.
This is more than five orders of magnitude smaller than
the unscreened Earth's magnetic field which is about half a Gauss at the surface
of the Earth.
Typical laboratory magnetic shields reduce the Earth's field by 2--3 orders of magnitude to a
mG level.

The order of magnitude estimate of the Zeeman splitting for \nhhh, where
magnetic interaction is dominated by 3 protons since 
the nitrogen $g$-factor is small ($\approx 0.4$), can be obtained as follows.
For a single proton, we have
\begin{equation}
\delta\omega_\mathrm{inv} = \mu_n g_p I_z  B \, ,
\label{zee1}
\end{equation}
where $\mu_n$ is the nuclear magneton, $g_p=5.6$ is the proton g-factor, $I_z$ is the proton
spin and $B$ is the strength of the magnetic field. 
Roughly, the maximum shift is realized when all
three proton spins are aligned (ortho-\nhhh), so we can use the above equation and put
$I_z=I_{z,1}+I_{z,2}+I_{z,3}=3/2$. 
With $\mu_n = 5.05\times10^{-24}$ erg~G$^{-1}$ and $B = 10^{-3}$ G, we obtain 
$\delta\omega_\mathrm{inv} \approx 6$ Hz.
Since the $(J,K)= (1,1)$ line belongs to the para-species of the ammonia
molecule, where the proton spins are not aligned, the expected frequency shift is
even lower for the transition observed by us.

One should also keep in mind that the Zeeman shifts turn to zero after averaging over magnetic 
quantum number $M$. 
Therefore, in the first approximation, magnetic fields 
induce a line broadening, but not a frequency shift. 
Thus, the systematic shift of the ammonia inversion transition of about $\sim 2$ kHz
detected in the starless molecular cores cannot be explained by a static
magnetic field effect. 

We conclude that static fields and thermal background
radiation produce systematic velocity shifts 
much smaller than 1 \ms. Thus, field effects do not contribute significantly to the systematic 
error of our observations.

\section{Conclusions}
\label{sect-6}

The present paper continues our studies where we use the ammonia inversion
transition in conjunction with low-lying rotational transitions
of other molecules to probe the dependence of
the electron-to-proton mass ratio $\mu$ on the ambient matter density.
As experimental tests, we suggest to observe dense prestellar
molecular cores located in the disk of the Milky Way.
Here we report on mapping of four molecular cores (L1498, L1512, L1517B, and L1400K)
in two molecular transitions of \nhhh\ $(J,K) = (1,1)$ and \hcccn\ $J = 2-1$
with high-spectral resolution at the 100-m Effelsberg telescope.
The main results are as follows.
\begin{enumerate}
\item[1.]
The completely resolved hfs components of the \nhhh\ (1,1) and \hcccn\ (2--1) transitions
allow us to determine the line centers with the precision of $\sim 1$ \ms,
which is comparable with laboratory uncertainties of the rest frequencies
of these transitions.
\item[2.]
In two cores, L1498 and L1512, with lowest Doppler noise,
we obtain a statistically significant positive velocity offset between
the rotational \hcccn\ and inversion \nhhh\ transitions of $26.9\pm1.2_{\rm stat}\pm3.0_{\rm sys}$ \ms.
In two other cores, L1517B and L1400K, the velocity offsets are, respectively, higher
and lower, but these deflections can be explained by the observed kinematic
structure in L1517B and L1400K.
If we assume that the measured velocity offset is caused by the electron-to-proton
mass ratio variation, then
\dmm\ = $26\pm1_{\rm stat}\pm3_{\rm sys}$ ppb.
The non-zero $\Delta\mu$
implies that at deep interstellar vacuum the
electron-to-proton mass ratio increases by
$\sim 3\times10^{-8}$ as compared with its terrestrial value.
\item[3.]
The reproducibility of the velocity offset at the same facility (Effelsberg telescope) on the
year-to-year base is very good, except for the L1400K target where the central point is
probably an outlier.
\item[4.]
We show that the effects of static electric and magnetic fields and the BBR-induced
Stark shifts are not larger than 1 \ms, and can be neglected in the total error budget.
\item[5.]
The results obtained tentatively support the hypothesis that in low-density environment
the equivalence principle may break. This may be a consequence of the chameleon-like scalar field. 
New laboratory measurements of the rest frequencies and new observations involving other
rotational transitions and other targets (e.g., Levshakov \etal\ 2010b)
are required to reach more definite conclusions.
\end{enumerate}

\begin{acknowledgements}
We are grateful to the staff of the Effelsberg 
radio observatory for assistance in our observations
and to Drs. Peter M\"uller, J\"urgen Neidh\"ofer, and Johann Schramml
for technical details on the 100-m telescope instrumentation.
The project has been supported in part by
DFG Sonderforschungsbereich SFB 676 Teilprojekt C4,
the RFBR grants No. 09-02-12223, 09-02-00352, and 08-02-92001,
the Federal Agency for Science and Innovations grant NSh-3769.2010.2,
the Program IV.12/2.5 of the Physical Department of the RAS,
and by the Chinese Academy of Sciences visiting professorship
for senior international scientists grant No. 2009J2-6.
\end{acknowledgements}

\end{document}